\documentclass[twocolumn,aps,prb,superscriptaddress,amsmath,amssymb,superscriptaddress]{revtex4-1}
\newcommand{\pagenumbaa}{1}
\usepackage{graphicx}
\usepackage{color}
\usepackage{braket}

\usepackage{verbatim}
\usepackage{amsmath}
\usepackage{mathtools}

\usepackage{braket}
\usepackage{float}
\usepackage[hypertexnames=false]{hyperref}
\usepackage{lipsum}
\usepackage{tabularx,ragged2e}
\newcolumntype{C}{>{\Centering\arraybackslash}X} 

\begin{document}

\title{Quantum Interference of Identical Photons from Remote GaAs Quantum Dots}

\author{Liang Zhai}
\email{liang.zhai@unibas.ch}
\affiliation{Department of Physics, University of Basel, Klingelbergstrasse 82, CH-4056 Basel, Switzerland}
\author{Giang N. Nguyen}
\affiliation{Department of Physics, University of Basel, Klingelbergstrasse 82, CH-4056 Basel, Switzerland}
\author{Clemens Spinnler}
\affiliation{Department of Physics, University of Basel, Klingelbergstrasse 82, CH-4056 Basel, Switzerland}
\author{Julian Ritzmann}
\affiliation{Lehrstuhl f\"ur Angewandte Festk\"orperphysik, Ruhr-Universit\"at Bochum, DE-44780 Bochum, Germany}
\author{Matthias C. L\"obl}
\affiliation{Department of Physics, University of Basel, Klingelbergstrasse 82, CH-4056 Basel, Switzerland}
\author{Andreas D. Wieck}
\affiliation{Lehrstuhl f\"ur Angewandte Festk\"orperphysik, Ruhr-Universit\"at Bochum, DE-44780 Bochum, Germany}
\author{Arne Ludwig}
\affiliation{Lehrstuhl f\"ur Angewandte Festk\"orperphysik, Ruhr-Universit\"at Bochum, DE-44780 Bochum, Germany}
\author{Alisa Javadi}
\affiliation{Department of Physics, University of Basel, Klingelbergstrasse 82, CH-4056 Basel, Switzerland}
\author{Richard J. Warburton}
\affiliation{Department of Physics, University of Basel, Klingelbergstrasse 82, CH-4056 Basel, Switzerland}

\begin{abstract}
Photonic quantum technology provides a viable route to quantum communication\cite{Sangouard2011,Yin2016}, quantum simulation\cite{Wang2019a}, and quantum information processing\cite{Qiang2018}. Recent progress has seen the realisation of boson sampling using 20 single-photons\cite{Wang2019a} and quantum key distribution over hundreds of kilometres\cite{Yin2016}. Scaling the complexity requires architectures containing multiple photon-sources, photon-counters, and a large number of indistinguishable single photons. Semiconductor quantum dots are bright and fast sources of coherent single-photons\cite{Strauf2007,Senellart2017,Liu2018,Uppu2021,Tomm2021}. For applications, a significant roadblock is the poor quantum coherence upon interfering single photons created by independent quantum dots\cite{Reindl2017,Weber2018}. Here, we demonstrate two-photon interference with near-unity visibility ($93.0\pm0.8$)\% using photons from two completely separate GaAs quantum dots. The experiment retains all the emission into the zero-phonon-line -- only the weak phonon-sideband is rejected -- and temporal post-selection is not employed. Exploiting the quantum interference, we demonstrate a photonic controlled-not circuit and an entanglement with fidelity ($85.0\pm 1.0$)\% between photons of different origins. The two-photon interference visibility is high enough that the entanglement fidelity is well above the classical threshold. The high mutual-coherence of the photons stems from high-quality materials, a diode-structure, and the relatively large quantum dot size. Our results establish a platform, GaAs QDs, for creating coherent single photons in a scalable way. 

\end{abstract}

\maketitle

\setcounter{page}{\pagenumbaa}
\thispagestyle{plain}

From large-scale quantum simulations\cite{Wang2019a} to multi-user quantum networks\cite{Sangouard2011}, the required number of photons soars as the complexity grows. Single-photon sources meeting these scaling needs do not yet exist. While current proof-of-principle demonstrations of photonic quantum applications rely mostly on parametric down-conversion sources\cite{Qiang2018,Llewellyn2020}, the adoption of deterministic sources is a clear trend\cite{Strauf2007,He2013_2,Senellart2017,Liu2018,Uppu2021,Basset2021,Tomm2021}: semiconductor quantum dots (QDs) are on-demand emitters of single photons with a significantly higher efficiency and photon generation-rate than down-conversion sources\cite{Senellart2017,Tomm2021}. In addition, QDs can be easily integrated into various nanostructures\cite{Liu2018,Grim2019,Uppu2021}. For single-photon generation, these advantages make QD-based sources arguably the best choice\cite{Senellart2017, Tomm2021}. However, to create a large number of photons, the prevalent approach -- active demultiplexing from a single QD\cite{Wang2019a} -- is not optimal. It introduces additional losses and leads to a large resource overhead, limiting the maximal number of photons.

A more advantageous approach is to create indistinguishable photons simultaneously from multiple QDs. This facilitates scaling up to higher photon numbers without sacrificing efficiency and photon generation-rate. It is an enabling technology with immediate use in photon-based boson sampling\cite{Wang2019a}, device-independent quantum key distribution\cite{Kolodynski2020} (QKD) and a photonic approach to measurement-based quantum computing\cite{Uppu2021} (MBQC). All these concepts call for quantum interference (i.e.\ interference with unity visibility) between photons from separate sources with no additional filtering loss. However, achieving such quantum interference using separate QDs has been a challenge for many years\cite{Patel2010,He2013,Giesz2015,Reindl2017,Weber2018,Zopf2018,You2021}. The reason is that the quantum interference is sensitive to the total noise in two uncorrelated solid-state environments. So far, the highest visibility on interfering photons directly created by two separate QDs (without filtering) is just above 50\% \cite{Reindl2017}.

In our experiment, photons from droplet-etched GaAs QDs are interfered and entangled [Fig.\ \ref{fig:oneHOM}(a)]. The QDs are hosted in $n$-$i$-$p$-diode heterostructures\cite{Zhai2020}. Three typical dots, QD1 -- QD3 are presented in Extended Data Fig.\ \ref{fig:plrf} and \ref{fig:l_l}.

For on-demand single-photon generation, we excite the QD resonantly with short laser pulses (duration $6\ \mathrm{ps}$). A $22\ \mathrm{GHz}$-bandwidth spectral filter is inserted in the collection to remove weak (4\%) phonon-sideband emission. Figure\ \ref{fig:oneHOM}(b) shows Rabi oscillations up to $6\pi$. In the following experiments, a $\pi$-pulse power is used, maximising the excited-state population to near-unity before photon creation. The purity of the photons, as characterised by $1-g^{(2)}(0)$, is $(99.0\pm 0.1)\%$. To probe the indistinguishability of single-QD photons, we perform a Hong-Ou-Mandel (HOM) experiment (see Methods). Figure\ \ref{fig:oneHOM}(c) shows the HOM measurements on QD2 $\mathrm{X}^{-}$ between consecutively emitted photons ($\mathcal{N} = 1$), i.e.\ the temporal separation between the photons corresponding to one repetition pulse-period ($\mathcal{D} = 13\ \mathrm{ns}$). The raw HOM visibility, defined as one minus the ratio of the central-peak intensity between co-polarised (HOM $\parallel$) and cross-polarised (HOM $\perp$) configurations, is $\mathcal{V}^{13\text{ns}}_{\text{raw}} = (95.8\pm1.2)\%$. The true HOM visibility for QD2 is $\mathcal{V}^{13\text{ns}} = (98.2\pm1.3)\%$ on correcting\cite{Santori2002} the raw visibility for the finite $g^{(2)}(0)$ and experimental imperfections (see Methods). Here, we calculate $\mathcal{V}$ by summing over the whole pulse period. This evaluation time-window ($T_{\mathrm{bin}} = 13\ \mathrm{ns}$) is significantly larger than the QD lifetime ($1/\Gamma_2 = 256\ \mathrm{ps}$), thus introduces no temporal post-selection.

The ambient environment of a QD can be static on a short time-scale in which case it has little impact on the coherence between consecutively emitted photons. However, slowly varying noise, for instance spectral fluctuations, decrease the HOM visibility on longer time-scales\cite{Wang2016,Thoma2016}. By adding a 200-metre fibre, we extend the temporal delay to $\mathcal{D} \sim1\ \mu \text{s}$, overlapping two photons separated by 77 pulse periods ($\mathcal{N} = 77$). The HOM interference of $13\ \text{ns}$ and $1\ \mu \text{s}$ separation are compared in Fig.\ \ref{fig:oneHOM}(d). We extract the $1\ \mu \text{s}$ visibility for QD2 ($T_{\mathrm{bin}} = 13\ \mathrm{ns}$), $\mathcal{V}^{1\mathrm{\mu \text{s}}}_{\text{raw}} = (95.7\pm2.0)\%$ and $\mathcal{V}^{1\mathrm{\mu \text{s}}} = (98.7^{+1.3}_{-2.0})\%$. 
As for the $13\ \mathrm{ns}$ case, the $1\ \mu \text{s}$ HOM visibility remains near-unity. As far as the HOM is concerned, the QD environment is static over a time-scale at least $\sim 4\times 10^3$ times larger than the QD's lifetime. [Comparable results are achieved with QD1: $1-g^{(2)}(0) = (98.7\pm0.2)\%$, $\mathcal{V}^{13\text{ns}} = (97.8\pm1.8)\%$, $\mathcal{V}^{1\mathrm{\mu s}} = (99.0^{+1.0}_{-1.8})\%$.] 

We now turn to the quantum interference between photons from independent sources [Fig.\ \ref{fig:twoHOM}(a)]. QD1 and QD2 are matched in both emission frequency and radiative rate (Methods). They are located in two individual wafer pieces and are hosted in two cryostats separated by 20 metres (in fibre length). Thus, the environments of QD1 and QD2 are completely uncorrelated. As such, the remote-interference visibility is sensitive to the noise in both semiconductor environments over a huge bandwidth, from 10$^{-4}$ Hz to 10$^{9}$ Hz. Despite this sensitivity to environmental noise, the visibility is near-unity, $\mathcal{V}^{\mathrm{remote}} = (93.0\pm 0.8)\%$ [Fig.\ \ref{fig:twoHOM}(b)]. The high visibility is not limited to only one pair of QDs: between QD1 and QD3 (another QD in the QD2-cryostat), a similar two-QD HOM visibility is observed, $\mathcal{V}^{\mathrm{remote}} = (92.7\pm 1.6)\%$.

The crucial aspects to achieve the high two-QD HOM visibility are an $n$-$i$-$p$ diode heterostructure\cite{Zhai2020} and well-controlled growth: the diode locks the QD charge-state via Coulomb blockade and suppresses spectral fluctuations; in the growth, impurities are minimised yielding ultra-pure materials. The GaAs QDs themselves are also important. GaAs QDs are larger in size ($\sim$100 nm in diameter and $5$-$7$ nm in height) with respect to InGaAs QDs. The larger QD-size increases the radiative rate thereby increasing the lifetime-limited linewidth, rendering the QD less sensitive to environmental fluctuations. Furthermore, the larger QD-size reduces slightly both the spin-noise, the main source of inhomogeneous broadening in low-charge-noise InGaAs QDs\cite{Kuhlmann2013}, and the exciton-phonon scattering rate, the main source of homogeneous broadening. 

On account of the uncorrelated environments, the two-QD HOM interference is equivalent to the coalescence of two photons from a single QD with an almost-infinite separation in time ($\mathcal{N}\rightarrow\infty$). Such coalescence is otherwise difficult to measure. For highly developed InGaAs QDs, 96$\%$ one-QD photon indistinguishability has been achieved\cite{Tomm2021} for temporal separation of up to 10$^{-6}$ s. Our two-QD HOM interference indicates that 93$\%$ one-QD indistinguishability can be achieved using gated GaAs QDs with a longer temporal separation -- as long as 10$^{4}$ s. As the residual noise in the QDs lies mostly at low-frequency ($<10^4$ Hz, Extended Data Fig.\ \ref{fig:nspectrum}), we expect to maintain a 98$\%$ one-QD indistinguishability even for a separation of 10$^{-4}$ s. 

Our two-QD HOM experiments are carried out under rigorous conditions: there is no Purcell enhancement of the radiative rate, no temporal post-selection, no narrow spectral filtering, and no active frequency stabilisation. Adding temporal post-selections or narrow spectral filters leads to higher two-QD HOM visibility, but the flux of usable photons goes down as the visibility goes up (Extended Data Fig.\ \ref{fig:timebin} and \ref{fig:filter}). For practical applications, it is clearly better to avoid this loss. In this way, connecting multiple QD sources does not introduce an additional loss to the source efficiency.

The small imperfections in the one-QD and two-QD HOM visibilities allow an estimation of the residual noise. The fast noise-process is exciton dephasing at rate $\Gamma^{*}$. This process is likely to arise from phonon scattering\cite{Thoma2016,Scholl2019}, resulting in a homogeneous linewidth broadening of $\Delta \nu_{\mathrm{H}}= \Gamma^{*}/\pi$. Random changes in the local environment of each QD are responsible for the slow noise-process\cite{Kuhlmann2013} -- a spectral fluctuation (inhomogeneous broadening). Assuming identical QDs and a Lorentzian probability distribution\cite{Kuhlmann2013} for each QD (with frequency width $\Delta \nu_{\mathrm{S}}$) to describe the spectral fluctuations, the two-QD HOM visibility $\mathcal{V}$ is given by:
\begin{equation}
    \mathcal{V}=\frac{1}{1+2\pi (\Delta \nu_{\mathrm{H}}+\Delta \nu_{\mathrm{S}})\cdot\tau_r},
    \label{Vtheory}
\end{equation}
where $\tau_r$ is the radiative lifetime. In the one-QD HOM experiment, the environment is static ($\Delta \nu_{\mathrm{S}} \rightarrow 0$) and the imperfection in the visibility determines the homogeneous broadening. We find $\Gamma^{*}=34\pm25\ \mathrm{MHz}$, equivalently $\Delta \nu_{\mathrm{H}}=11\pm8\ \mathrm{MHz}$. In the two-QD HOM experiment, $\mathcal{V}$ depends on both exciton dephasing and spectral fluctuations allowing $\Delta \nu_{\mathrm{S}}=34 \pm 15\ \mathrm{MHz}$ to be determined. The average single-QD linewidth estimated from this HOM-analysis, $1/(2\pi \tau_r)+ \Delta \nu_{\mathrm{H}} +\Delta \nu_{\mathrm{S}}$, matches well with the measured linewidth. This analysis shows that together, the one-QD and two-QD HOM experiments enable the fast and slow noise-processes to be determined separately (Supplementary Information). 

The interference visibility of photons from independent GaAs QDs ($\mathcal{V}^{\mathrm{remote}}$) is an important metric for the application of this system in quantum technologies. The visibility is comparable to that achieved in trapped ions\cite{Maunz2007,Stephenson2020} and cold atoms\cite{Beugnon2006}, the seemingly most identical emitters.
It is also comparable to the visibilities achieved with state-of-the-art parametric photon sources\cite{Llewellyn2020} and with coherent scattering from solid-state emitters\cite{Stockill2017}. However, both parametric sources and coherent scattering operate in an intrinsically probabilistic manner where the photon generation rate is compromised to achieve high interference visibility. The two-QD HOM visibility is slightly higher than that achieved with remote nitrogen-vacancy centres\cite{Bernien2013,Humphreys2018} for which temporal post-selections and narrow frequency-filters are required: both considerably decrease the impact of spectral fluctuations at the expense of a reduced efficiency.

To map out the dependence of the two-QD HOM visibility on the two-photon coalescence, we deliberately reduce the overlap either temporally or spectrally.
As the delay between the two QDs' photons $\delta\mathrm{t}$ increases, the interference visibility decreases exponentially [Fig.\ \ref{fig:twoHOM}(c)], a consequence of the reduced temporal overlap of the two wave-packets. This is in excellent agreement with theoretical calculations for exponentially-decaying wave-packets\cite{Kambs2018} (Supplementary Information). 
When reducing the spectral overlap, the two-QD HOM visibility follows a Lorentzian profile on detuning $\Delta$ [Fig.\ \ref{fig:twoHOM}(d)], again exactly as expected from theory. This configuration offers a further test of the photon coherence. When the two photons are slightly detuned in frequency, a quantum beat is expected in the time-dependence of the intensity correlation function\cite{Kambs2018}. These quantum beats are very clearly observed: Figure \ref{fig:twoHOM}(e,f) show quantum beats in the central HOM peak for $\Delta/\Gamma = 0.31$ and $\Delta/\Gamma = 0.87$, respectively. The oscillation period decreases as $\Delta$ increases. These pronounced quantum beats match nicely with theoretical calculations, and from a different perspective, reflect the mutual coherence of the photons created by the remote QDs.

\begin{figure*}[t]
\includegraphics[width=2\columnwidth]{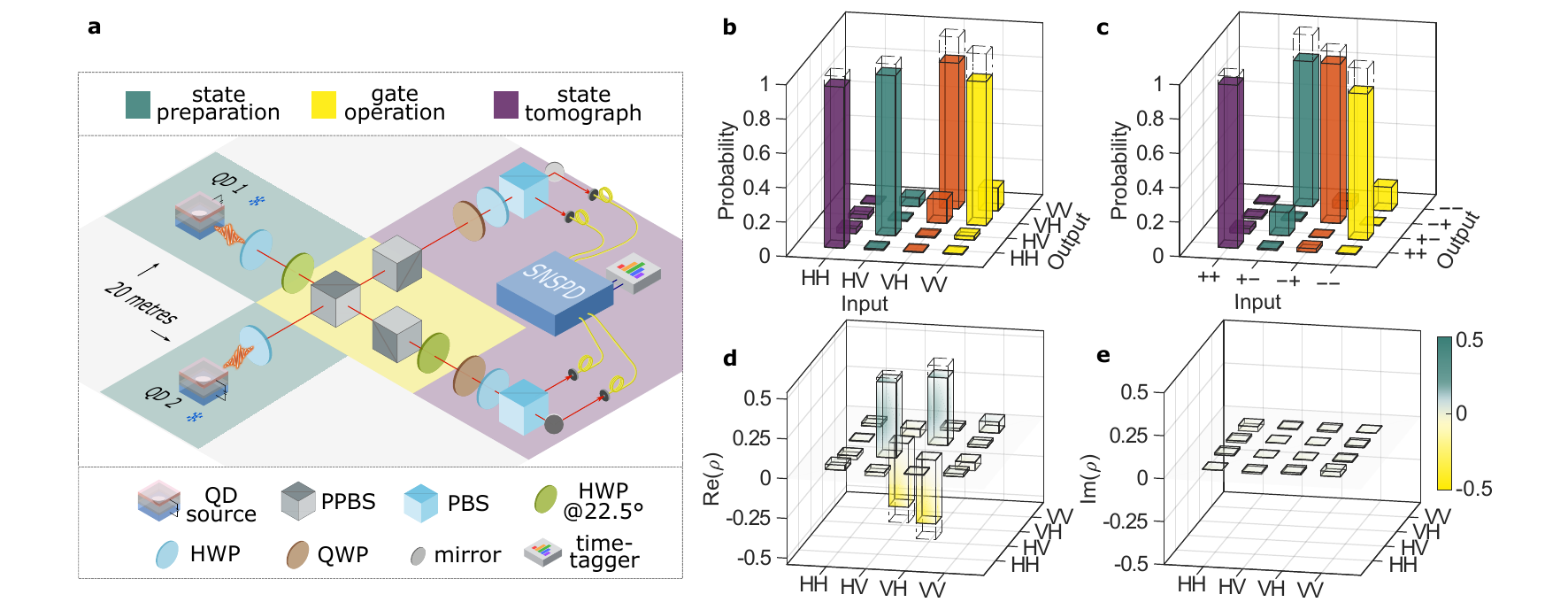}
\caption{\label{fig:cnot}\textbf{A quantum logic gate connecting photons from remote quantum dots.} \textbf{(a)} Sketch of the experimental setup for the CNOT gate based on two-photon interference between remote QDs. The photons from QD2 are used for the control qubit, and the photons from QD1 for the target qubit. For state preparation (highlighted in green), photons from the two QDs are carefully balanced to the same flux where their polarisations prepare the input states. The CNOT gate is highlighted in yellow. Half-wave plates (HWP) at 22.5$^\circ$ are Hadamard gates bringing $H$- and $V$-polarisations to the superposition. The central partially polarising beamsplitter (PPBS) transmits $H$-polarised light $\text{T}_{H}=1$, while partially reflects (transmits) $V$-polarised light $\text{R}_{V}=2/3$ ($\text{T}_{V}=1/3$). Upon simultaneous arrival at this PPBS, quantum interference between two indistinguishable photons provides a $\pi$-phase shift in the $\ket{VV}$ amplitude\cite{Kiesel2005}. The two PPBSs in opposite arms are each rotated by 90$^\circ$ with respect to the central one, such that their $\text{T}_{H}=1/3$, reducing the amplitude of $H$-polarised components to the same of $\ket{VV}$. Together, the three PPBSs constitute a controlled-Z gate. After the gate operation, the quantum state is projected in a quantum tomography setup (highlighted in purple). For every input state, the output state is mapped out using four simultaneous coincidence-measurements. \textbf{(b,c)} The truth tables for $\ket{H}/\ket{V}$ and $\ket{+}/\ket{-}$ bases. The coincidence events are converted to probabilities by normalising to the respective input state. For each input state, we accumulate total coincidence counts of $\sim500$. As in the HOM analysis, we do not use any temporal post-selection, $T_{\mathrm{bin}} = 13\ \mathrm{ns}$. The empty dashed bars represent the ideal CNOT operation. \textbf{(d,e)} The real (d) and imaginary (e) parts of the density operator of state $\ket{\Psi^-}$ created by the CNOT gate. The real parts of the measured density matrix, Re$(\rho)$, are close to ideal (represented by the empty bars). In the imaginary parts, Im$(\rho)$, the intensities are all below $0.03$.}
\end{figure*}

Indistinguishable photons from distant QDs enable the creation of a CNOT gate. Quantum information can be independently encoded into single-photon streams created by separate QDs using polarisation: $\ket{H} = \ket{0}$, $\ket{V} = \ket{1}$. The optical CNOT circuit is realised by a combination of two half-wave plates (HWP) at $22.5^\circ$ and three partially polarising beamsplitters\cite{He2013_2} (PPBS), as depicted in Fig.\ \ref{fig:cnot}(a). Each $22.5^\circ$ HWP acts as a Hadamard gate; the three PPBSs constitute a controlled-phase (CZ) gate\cite{Kiesel2005}. The CZ gate relies on coincidence clicks on detectors in two opposite output arms (Supplementary Information). The gate operation is probabilistic with a success probability of 1/9. 

We evaluate the gate performance using the input-output relations in both the computational basis $\ket{H}/\ket{V}$ and the basis defined by the linear superpositions $\ket{\pm} = 1/\sqrt{2}(\ket{H}\pm\ket{V})$. Ideally, in the $\ket{H}/\ket{V}$ basis ($ZZ$ basis in Pauli matrix language) the target qubit flips the sign when the control reads logical one. In the $\ket{+}/\ket{-}$ basis ($XX$ basis), the target qubit decides whether its control counterpart undergoes a flip. Experimentally, the output states are analysed by a quantum tomography setup. The corresponding truth tables are shown in Fig.\ \ref{fig:cnot}(b,c). The CNOT operation fidelity in $ZZ$ and $XX$ bases are $\mathcal{F}_{ZZ} = (88.90\pm5.34)\%$, and $\mathcal{F}_{XX} = (89.34\pm5.29)\%$. We calculate the bound for the overall quantum process fidelity\cite{He2013_2} based on ($\mathcal{F}_{ZZ}+\mathcal{F}_{XX}-1) < \mathcal{F}_{\mathrm{proc}} < \mathrm{min}(\mathcal{F}_{ZZ},\mathcal{F}_{XX})$, yielding $(78.24\pm7.53)\%< \mathcal{F}_{\mathrm{proc}} < (88.90\pm5.34)\%$.

Finally, we demonstrate the ability to create maximally entangled states using photonic qubits from remote QDs. This is the hallmark of the CNOT operation: by preparing the input state as $\ket{-}_{\text{c}}\ket{V}_{\text{t}}$, the Bell state $\ket{\Psi^-} = \frac{1}{\sqrt{2}}(\ket{HV}-\ket{VH})$ is produced. To characterise fully the produced state, quantum state tomography is performed with a series of 36 coincidence measurements, Extended Data Fig.\ \ref{fig:sv_cnot}, followed by state reconstruction using a maximum-likelihood-estimation algorithm\cite{James2001}. The real and imaginary parts of the reconstructed density matrix $\rho$ are shown in Fig.\ \ref{fig:cnot}(d,e). We obtain an entanglement fidelity of $\mathcal{F}_{\mathrm{\Psi}^-} = (85.02\pm0.97)\%$, which is well above the threshold of $(2+3\sqrt{2})/8= 0.78$ for violating Bell inequalities\cite{Kiesel2005}. To quantify the entanglement, we calculate the concurrence $C = (74.67\pm 1.93)\%$ and the linear entropy $S_{\mathrm{L}} = (34.04\pm 1.94)\%$. These values indicate a high degree of entanglement established using two separate streams of photons.

Based on the measured two-QD HOM visibility, $\mathcal{V}^{\mathrm{remote}} = 93.0\%$, the expected entanglement fidelity is 90.2$\%$ (Methods); the expected process fidelity in the computational basis is 93.9$\%$ (Supplementary Information). The slight mismatch with respect to the experiment is likely due to imperfections in the optical elements and the non-zero $g^{(2)}(0)$ values. The CNOT demonstration highlights the importance of high-visibility two-QD HOM-interference: if the visibility is low, the entanglement fidelity will be worse; a minimal visibility of $\mathcal{V} = $83\% is required in our scheme for the generated entanglement fidelity to surpass the Bell-inequality violation threshold, $\mathcal{F}_{\mathrm{\Psi}^-}> 78\%$.

In conclusion, we have demonstrated that GaAs QDs are interconnectable sources of indistinguishable single photons. The near-unity mutual-coherence between photons created by separate QDs points to the potential of employing multiple QD-sources in quantum applications. The extraction efficiency can be boosted by the Purcell effect upon coupling the QDs to a single optical mode\cite{ Uppu2021,Tomm2021,Senellart2017}. The HOM visibility should also benefit from the reduced lifetime (Purcell effect, Eq.\ \ref{Vtheory}): with the present noise level, a Purcell factor $F_{\mathrm{p}} \sim 10$ should result in one-QD and two-QD HOM visibilities of $99.6\%$ and $99.0\%$, respectively. From a quantum-information perspective, increasing the number of identical photons to $\sim50$ will lead to quantum advantage in a boson sampling experiment\cite{Uppu2021}. This number of photons is within reach, for instance using several gated GaAs QDs together with current photonic technologies\cite{Wang2019a}. For MBQC, once few-photon cluster states are generated by individual QDs\cite{Istrati2020,Cogan2021}, projective entangling gates allow the small clusters to be ``fused" into large-scale computational resources\cite{Uppu2021}. High remote-interference visibility points to a high-fidelity entangling operation. From a quantum communication perspective, highly indistinguishable photons forge a coherent link between remote QDs, a route to the realisation of device-independent QKD with high key rates\cite{Kolodynski2020}. Moreover, GaAs QDs can be brought into exact resonance with rubidium transitions, allowing the storage of QD photons in a rubidium-based quantum memory\cite{Wolters2017}. 

\section*{Acknowledgements}\vspace{-3mm}
We thank Peter Lodahl, Doris Reiter, Philipp Treutlein, and Ravitej Uppu for stimulating discussions. The work was supported by NCCR QSIT and SNF Project No.s\ 200020$\_$175748 and 200020$\_$204069. LZ, GNN, AJ received funding from the European Union’s Horizon 2020 Research and Innovation Programme under the Marie Skłodowska-Curie grant agreement No.\ 721394 (4PHOTON), No.\ 861097 (QUDOT-TECH), and No.\ 840453 (HiFig), respectively. JR, ADW and AL acknowledge financial support from the grants DFH/UFA CDFA05-06, DFG TRR160, DFG project 383065199, and BMBF QR.X Project 16KISQ009.

\section*{Author Contributions}\vspace{-3mm}
LZ, GNN, CS, AJ carried out the experiments. JR, LZ, MCL, ADW, AL designed and grew the sample. CS, GNN, LZ fabricated the sample. LZ, GNN, CS, AJ, MCL, RJW analysed the data. LZ, GNN, CS, RJW wrote the manuscript with input from all the authors.
\section*{Competing Interests}\vspace{-3mm}
The authors declare no competing interests.
\clearpage

\begin{figure*}[t]
\centering
\includegraphics[width=2\columnwidth]{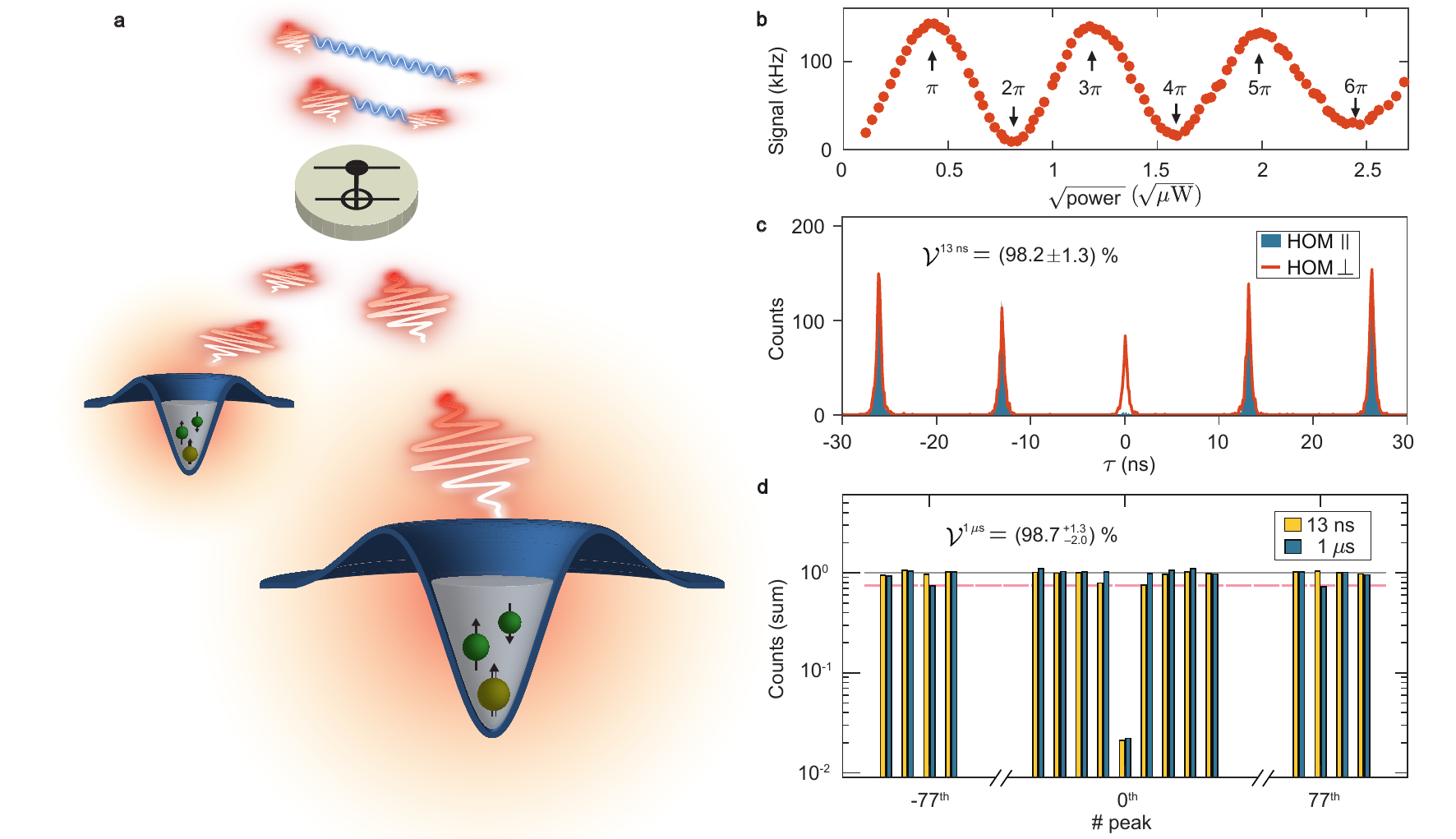}
\caption{\label{fig:oneHOM}\textbf{Coherent single photons from GaAs quantum dots at a wavelength of 780 nm.} \textbf{(a)} A schematic view of an entangling gate between photons from two separate GaAs QDs. Each QD is excited independently (creating two electrons and one hole, the $\mathrm{X}^{-}$) and generates a stream of single photons. When the photons are identical with each other in separate streams, quantum interference takes place in the gate, entangling the two photon-streams. \textbf{(b)} Resonance fluorescence from an individual QD (QD2, $\mathrm{X}^{-}$) reveals Rabi oscillations. The first Rabi cycle shows the highest contrast -- at a Rabi power of $2\pi$, the signal intensity drops to just $\sim6\%$ of that at $\pi$. \textbf{(c)} Hong-Ou-Mandel (HOM) experiments on QD2 $\mathrm{X}^{-}$ showing quantum interference of photons created $13\ \mathrm{ns}$ apart. By preparing the two photons in the same polarisation (co-polarisation, HOM $\parallel$) the central peak almost vanishes due to quantum interference. On the contrary, the photons become distinguishable in cross-polarisation (HOM $\perp$): the quantum interference no longer takes place and the central peak appears. \textbf{(d)} Two-photon interference for two different delays, $13\ \mathrm{ns}$ and $1\ \mathrm{\mu s}$. Each bar represents the total coincidence counts summing over a whole pulse period. The coincidence probability of the central peak is nearly identical for the two delays, while the side peaks are flat and close to one (grey line). The red line indicates the decrease to $75\%$ due to route probability. The uncertainties in the interference visibilities ($\mathcal{V}^{13\text{ns}}$ and $\mathcal{V}^{1\mathrm{\mu \text{s}}}$) are the 1$\sigma$ random error, arising from the measurement setup and the shot noise of the detectors (Methods).}
\end{figure*}

\begin{figure*}[t]
\centering
\includegraphics[width=2\columnwidth]{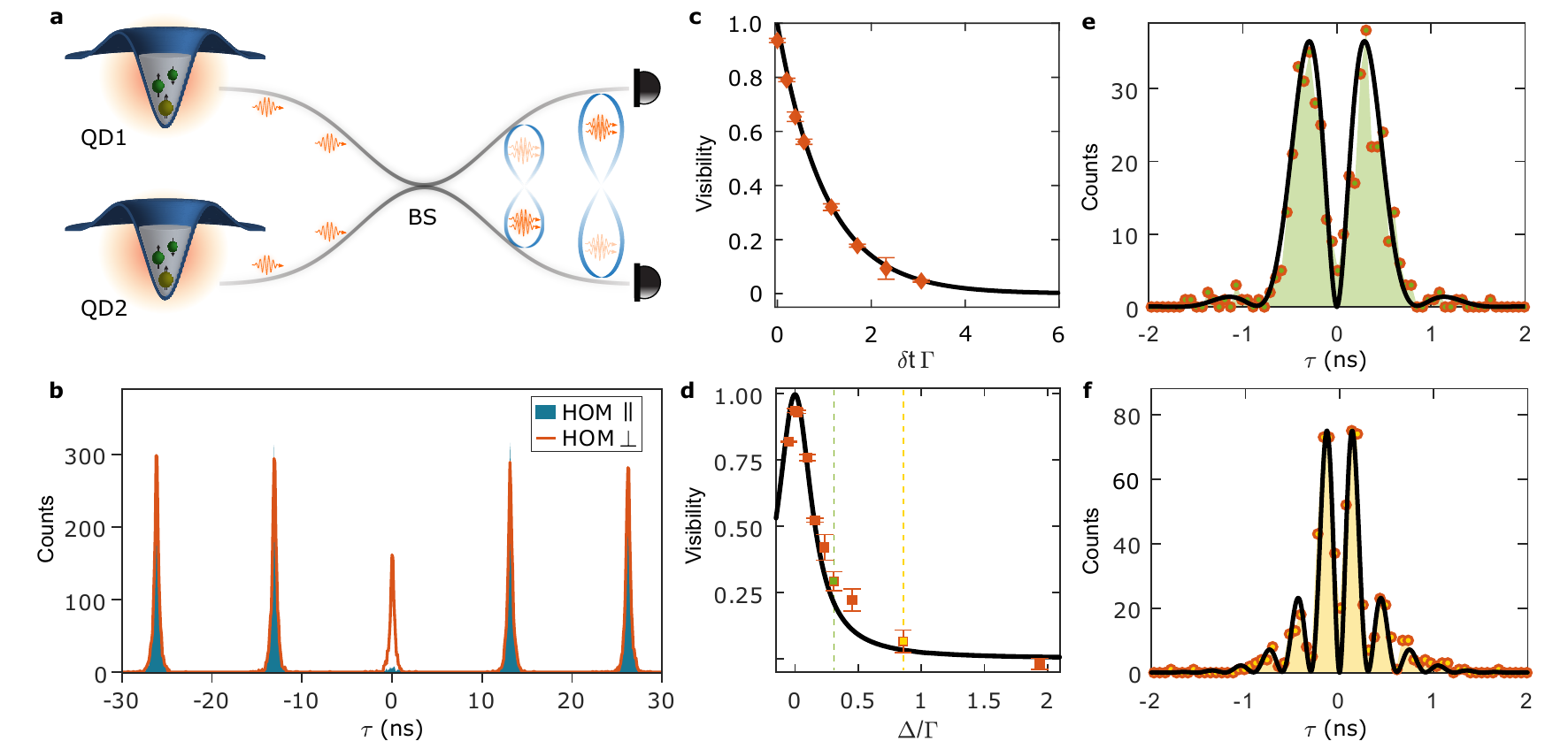}
\caption{\label{fig:twoHOM}\textbf{Two-photon interference from remote quantum dots.} \textbf{(a)} Schematic representation of the quantum interference between photons from two remote GaAs QDs. Provided the photons are indistinguishable, they coalesce and exit the beamsplitter (BS) at the same port. \textbf{(b)} HOM interference from photons generated by two remote QDs. The vanishing central peak shows a raw HOM visibility of $(90.9\pm0.8)\%$ and a true two-photon interference visibility of $(93.0\pm 0.8)\%$. \textbf{(c,d)} Two-QD HOM visibility as a function of normalised temporal delay $\mathrm{\delta t}$ and spectral detuning $\mathrm{\Delta}$ between the two QDs. In (c), the QD2 photons are delayed with respect to QD1 by a delay $\delta\mathrm{t}$. In (d), the frequency detuning $\Delta$ between the two QDs is varied by exploiting the exquisite frequency control provided by the electrical gates. The uncertainties of interference visibilities in (b-d) represent the 1$\sigma$ error (Methods). \textbf{(e,f)} Pronounced quantum beats are revealed in the two-QD HOM central peak when the QDs are slightly detuned in frequency. The green and yellow colours indicate the detuning in (d). The solid-lines in (c-f) show the quantum-optical theory. In the theory, we assume the wave-packets of remote QDs are identical except for a small delay in (c) and a small detuning in (d-f).}
\end{figure*}

\section*{Methods}\vspace{-3mm}

GaAs QDs create single photons at deep-red wavelengths (750 -- 800 nm), a very convenient spectral band: low-loss optical fibres, semiconductor lasers, and highly efficient single-photon detectors are readily available. The QDs studied in this paper are grown in a molecular beam epitaxy chamber\cite{Nguyen2020}. The sample consists of an $n$-$i$-$p$ diode heterostructure\cite{Zhai2020}. The QDs are formed on an intrinsic $\mathrm{Al}_{0.33}\mathrm{Ga}_{0.67}\mathrm{As}$ matrix using a local droplet etching technique\cite{Gurioli2019,Heyn2009}. For the $n$- and $p$-doped layers, $\mathrm{Al}_{0.15}\mathrm{Ga}_{0.85}\mathrm{As}$ instead of $\mathrm{Al}_{0.33}\mathrm{Ga}_{0.67}\mathrm{As}$ is employed to avoid the formation of DX centres\cite{Mooney1990}. The detailed heterostructure design and the growth parameters are identical to those in Ref.\ \onlinecite{Zhai2020}.

\subsection{Optical properties of GaAs QDs}\label{optical}
The $n$-$i$-$p$ diode enables deterministic control of the QD charge via Coulomb blockade\cite{Warburton2013}. An example is shown in Extended Data Fig.\ \ref{fig:plrf}(a) for QD2. The photoluminescence undergoes abrupt jumps, i.e.\ charge-states appear one after another as the gate voltage increases. The heterostructure design also allows large-range frequency tuning within one charge plateau via the quantum-confined Stark effect. The tuning range $R$ is typically\cite{Zhai2020} above $200\ \mathrm{GHz}$, much larger than the QD linewidths. This is exploited in the resonance fluorescence charge-plateaus on $\mathrm{X}^{-}$ of QD1 and QD2 [Extended Data Fig.\ \ref{fig:plrf}(b,c)], where the precise adjustments of external gate voltages bring the two QDs into resonance. Comparing the Stark tuning range $R$ with the full-width-at-half-maximum of the emission ensemble $W$ ($W = 11\ \mathrm{nm}$, equivalently $5.4\ \mathrm{THz}$), we find $R/W\simeq3.6\%$. This ratio is larger than the equivalent tuning capability of InGaAs QDs; while the emission ensemble bandwidth $W$ is smaller than the typical InGaAs one. This large-range frequency tuning spares the tedious searching process for frequency-matched emitters.

The GaAs QDs in the $n$-$i$-$p$ diode system exhibit very little noise under continuous-wave (CW) excitation. This is revealed by their near-lifetime-limited linewidths. In Extended Data Fig.\ \ref{fig:l_l}, we show lifetime- and linewidth-measurements for QD\ $1,2,3$ and their associated data fittings. The extracted decay rates for the three QDs are $\Gamma_1 = 3.75\ \mathrm{GHz}$, $\Gamma_2 = 3.91\ \mathrm{GHz}$ and $\Gamma_3 = 3.54\ \mathrm{GHz}$, respectively. This corresponds to lifetime-limited optical linewidths of $\Gamma_1/2\pi = 597\ \mathrm{MHz}$, $\Gamma_2/2\pi = 623\ \mathrm{MHz}$ and $\Gamma_3/2\pi = 564\ \mathrm{MHz}$. The measured linewidths for QD 1 -- 3 are just $8.4\%$, $9.6\%$, and $14.0\%$ above the lifetime limit. 

Photons from the GaAs QDs are collected by an aspheric lens (NA = 0.71) and a dark-field microscope. The dark-field microscope relies on a cross-polarisation scheme\cite{Kuhlmann2013a} to separate the QD photons from the excitation laser. There is an intrinsic loss of 50$\%$. The microscope has an overall efficiency of 15$\%$ - 20$\%$. The setup collection efficiency is characterised under resonant pulsed excitation. The QD photons are counted by an avalanche photodiode ($\sim70\%$ detection efficiency at 780 nm). The connection loss on the way to the avalanche photodiode (APD) is around 1 dB. At the APD, we detect 150 kHz - 200 kHz photons when exciting the QD with 76.3 MHz laser pulses. This corresponds to a collection efficiency of 0.3$\%$ – 0.4$\%$ from emitter to detector. Taking into account the setup imperfections, QD photons are collected by the first lens with an efficiency of $\sim2.3\%$.

\subsection{Auto-correlation measurements under continuous-wave excitation}\label{autocc}

The $n$-$i$-$p$ diode helps to stabilise the photon emission from GaAs QDs, avoiding the ``blinking" issue\cite{Zhai2020}. Blinking, random telegraph noise in the QD emission intensity, results in a bunching characteristic in an auto-correlation measurement. In our system, when the QDs are excited resonantly with a narrow-bandwidth CW laser, the auto-correlation ($g^{(2)}$) remains flat and close to unity [Extended Data Fig.\ \ref{fig:nspectrum}(a)]. The blinking is not present at the resonance condition.

The CW auto-correlation measurement also allows the noise in the local environment of a QD to be studied. We obtain the noise power density spectrum of QD1 by performing a fast Fourier transform on $g^{(2)}(\tau)$ [Extended Data Fig.\ \ref{fig:nspectrum}(b)]. The noise spectrum shows that the QD is influenced by very slow noise, i.e.\ noise at low frequencies. The noise becomes less as frequency f increases; it becomes mostly flat from f $>10^{4}$ Hz up to f $\sim10^{6}$ Hz (not shown in the figure). The frequency dependence of the noise can be fit to a Lorentzian function very well, as shown in Extended Data Fig.\ \ref{fig:nspectrum}(b). The noise at f $>10^{4}$ Hz is very low compared to the low-frequency part. 

By moving the CW laser frequency to the flank of the QD spectrum (detuning $\Delta = \Gamma/2$), the sensitivity to charge noise is enhanced in the auto-correlation measurement\cite{Kuhlmann2013}. The corresponding noise spectrum is shown as the light-blue curve in Extended Data Fig.\ \ref{fig:nspectrum}(c). The noise at f $<10^{2}$ Hz is higher in this situation with respect to Extended Data Fig.\ \ref{fig:nspectrum}(b), such that the dominant contribution is the charge noise. The orange curve in the same plot corresponds to the situation when we selectively enhanced the spin noise\cite{Kuhlmann2013}. In this case, the noise at $10^2 - 10^{4}$ Hz is increased, while the f $<10^{2}$ Hz part is reduced. This is likely to be the spin noise. This measurement infers that the spin noise is slightly faster than the charge noise in the GaAs QDs.

\subsection{Finding two similar quantum dots}

To achieve the quantum interference between photons created by separate QDs, the QDs should be matched in both frequency and radiative rate. In our experiment, QD1 and QD2 are tuned into resonance by the Stark effect (Extended Data Fig.\ \ref{fig:plrf}). Their radiative rates are naturally well-matched: the temporal overlap is $\Gamma_1/\Gamma_2 =96\%$, while the spectral overlap is $\Delta \nu_{1}/\Delta \nu_{2} = 95\%$ (with $\Delta\nu_{1,2}$ denoting the measured linewidths). 

We describe the efforts involved in finding two QDs with good frequency and decay-rate overlap. To assess this, we deliberately fix QD1 as one of the candidates in one cryostat (cryo1) and look for another candidate in the second cryostat (cryo2). Two parameters are critical for the search -- the $\mathrm{X}^{-}$ emission frequency and its radiative lifetime. Employing spatially resolved photoluminescence mapping\cite{Lobl2019} (PL-map), we establish the connection between QD emission frequency and QD spatial location in a $25\times25\ \mathrm{\mu m}^2$ region. A PL-map takes usually $5-6$ hours to record and contains $\sim 200$ QDs, with all QD positions and emission frequencies logged in a coordinate system. Among the QDs in one PL-map, we find typically $3-6$ QDs whose $\mathrm{X}^{-}$ frequency is close to QD1. The position information allows us to move to these selected QDs one by one and investigate their linewidths, lifetimes, and frequency tuning ranges, similar to the measurements in Extended Data Fig\ \ref{fig:l_l}. The lifetimes of QDs in our sample are similar: based on the lifetime measurements of ten randomly chosen QDs, the coefficient of variation is only $C_v(\tau) = 0.2$. Thanks to this small spread in lifetime and the near-lifetime-limited optical linewidths, screening a QD suitable in both lifetime and frequency is not demanding. For instance, it took two PL-maps to find QD2. This screening process can be further simplified in the future if lifetime tuning is in place, e.g.\ by coupling to a microcavity\cite{Tomm2021}. Narrowing the QD emission ensemble via growth\cite{Keil2017} and adding a large-range frequency-tuning capability via external strain\cite{Zhai2020_2,Grim2019} can also speed up the process.

\subsection{Hong-Ou-Mandel experiments with one quantum dot}

In the single-QD HOM measurements, photons from a single QD are filtered by a broadband grating-based filter ($22\ \mathrm{GHz}$ bandwidth) and then sent into a Mach-Zehnder interferometer (Supplementary Information). A photon travelling through the longer path of the interferometer overlaps on a symmetric beamsplitter with the $\mathcal{N}^{\text{th}}$ subsequently emitted photon travelling through the shorter path. After the beamsplitter, the photons are counted by two superconducting nanowire single-photon detectors (SNSPDs) and analysed by a time-to-digital converter. Details regarding the experimental setup are described in Supplementary Note 1. Experimental results of one-QD HOM are presented in Fig.\ \ref{fig:oneHOM} and Supplementary Fig.\ 4, 5. The data, e.g.\ as shown in Fig.\ \ref{fig:oneHOM}(c), is integrated for 320 minutes with a count-rate of 2 kHz per detector (HOM $\parallel$). The bin size of the time-to-digital converter is set to 100 ps. Summing up the coincidence counts for each peak using a 13 ns time-window [HOM $\parallel$, Fig.\ \ref{fig:oneHOM}(c)], there are 20 counts for the central peak and 942 counts on average for the side peaks except for the first ones. For HOM $\perp$ configuration, we collect on average 1038 counts for the side peaks and 536 counts for the central peak. The raw one-QD HOM visibility is thus calculated by $\mathcal{V}^{13\ \mathrm{ns}}_{\mathrm{raw}} =1- (\frac{20}{942})/ (\frac{536}{1038})=95.8\%$ (for QD2).

To determine the true overlap $\mathcal{V}$ of the two single-photon states produced by the QD, we account for the finite $g^{(2)}(0)$ and for imperfections in the setup (the imperfect classical interference visibility and the small imbalance in the ``50:50" beamsplitter). For the finite $g^{(2)}(0)$ value (likely caused by a re-excitation process\cite{Fischer2017}), we assume that the two photons in the occasionally created $\ket{2}$-state are distinguishable\cite{Tomm2021}. For one-QD HOM measurements, we follow the calculations in Ref.\ \onlinecite{Santori2002} and arrive at: 
\begin{equation} \label{eq:m1}
    \mathcal{V} = \frac{1}{(1-\epsilon)^2}\Big(\frac{R^2+T^2}{2RT}\Big)\Big[1+2\cdot g^{(2)}(0)\Big]\ \mathcal{V}_{\mathrm{raw}}.
\end{equation}
Here, $R$ and $T$ are the reflection and transmission coefficients of the beamsplitter, and $(1-\epsilon)$ is the classical visibility of the interferometer. For the one-QD HOM setup (Supplementary Information), $R = 0.490$ and $T = 0.510$; the classical visibility $(1-\epsilon)$ is $0.998$ for $\mathcal{D} = 13\ \mathrm{ns}$, and $0.995$ for $\mathcal{D} = 1.01\ \mathrm{\mu s}$. 

The uncertainties on single-QD HOM visibilities (and also $g^{(2)}(0)$s) in the main text represent the errors arising from the instabilities of the measurement setup and the shot noise of the detectors. They are calculated by dividing the total coincidence events into smaller parts according to the integration time, and computing the statistical standard deviation. 

\subsection{Hong-Ou-Mandel experiments with separate quantum dots}

In the two-QD HOM measurements, single photons from independent QDs are adjusted to the same intensity and sent to a 50:50 beamsplitter. Experimental details are presented in Supplementary Note 2. Experimental results of two-QD HOM are shown in Fig.\ \ref{fig:twoHOM}(b) and Supplementary Fig.\ 8. The bin size of the time-to-digital converter is set to 100 ps. The data shown in Fig.\ \ref{fig:twoHOM}(b) is integrated for 145 minutes with a count-rate of on average 4 kHz per detector (HOM $\parallel$). The inverse of this integration time represents the lower bound of the noise sensitivity bandwidth, i.e.\ 10$^{-4}$ Hz. (The upper bound is set by the lifetime of the QDs, which is on the order of 10$^{9}$ Hz.) In total, there are 89 coincidence counts in the HOM central peak (no temporal post-selection), and on average 1832 counts for the side peaks. For HOM $\perp$, the count rates are similar. We integrate until the side peaks reach a similar intensity of the HOM $\parallel$ case. In total, there are 942 coincidence counts for the central peak and 1768 counts for the side peak. For presenting the data, the coincidence counts in the HOM $\perp$ configuration are scaled with respect to the HOM $\parallel$ case by a factor of $1832/1768 = 1.036$ such that the relative area of the HOM central peaks can be directly compared. The raw two-QD HOM visibility (for QD1/QD2) is thus calculated by $\mathcal{V}^{\mathrm{remote}}_{\mathrm{raw}} =1- (\frac{89}{1832})/(\frac{942}{1768})=90.9\%$. 

To calculate the true two-QD HOM visibilities, the major difference compared to the one-QD experiment is the absence of the Mach-Zehnder interferometer. The first beamsplitter in the Mach-Zehnder interferometer gives rise to the factor of two in front of the $g^{(2)}(0)$ in Eq.\ \ref{eq:m1}. Moreover, for the two-QD HOM, the $g^{(2)}(0)$s of both QDs are taken into account. We outline the derivation of the true two-QD HOM visibility $\mathcal{V}$ in Supplementary Note 2. We arrive at:
\begin{equation} \label{eq:m2}
    \mathcal{V} = \frac{1}{(1-\epsilon)^2}\ \Big(\frac{R^2+T^2}{2RT}\Big)\Big[1+\frac{1}{2}\Big(g^{(2)}_{\mathrm{QD_i}}(0)+g^{(2)}_{\mathrm{QD_j}}(0)\Big)\Big]\ \mathcal{V}_{\mathrm{raw}},
\end{equation}
with $i,j$ denoting the two QDs. For the two-QD HOM setup, $R = 0.498$ and $T = 0.502$; the classical visibility $(1-\epsilon)=0.996$.

The uncertainties on $\mathcal{V}^{\mathrm{remote}}$ in Fig.\ \ref{fig:twoHOM} (b-d) contain the errors arising from the shot noise of the detectors and uncertainties in the experimental setup. They are calculated in the same way with respect to that of single-QD HOM visibility. 

To model the HOM interference between the photons generated by remote QDs, we adopt and modify the analytical treatment developed in Ref.\ \onlinecite{Kambs2018}. Derivations are presented in Supplementary Note 2.A. Assuming a Lorentzian distribution to describe spectral fluctuation of each QD, the result for the two-QD HOM visibility is: 
\begin{equation} \label{eq:m3}
    \mathcal{V} =  \frac{2(\gamma+\Xi\pi)}{(\gamma+\Xi\pi)^2+4\pi^2\Delta^2}\cdot\frac{e^{-|\delta t|/\tau_e}}{(\tau_i+\tau_j)}.
\end{equation}
Here, $\delta t$ and $\Delta$ represent the temporal delay and spectral detuning, respectively, of one wave-packet with respect to the other. $\tau_{i,j}$ stands for the radiative lifetime of the two QDs, and $\tau_{e}$ is the lifetime of the ``early" photon ($e =i$ or $j$). $\gamma$ is the overall phase relaxation-rate, $\gamma = \gamma_i+\gamma_j$, where $\gamma_{i,j} = 1/(2\tau_{i,j}) + \Gamma^*_{i,j}$. $\Gamma^*_{i,j}$ and $\Xi$ are included to model the effects of phonon-induced pure dephasing and spectral fluctuation processes, respectively. In Eq.\ \ref{eq:m3}, we assume the total spectral fluctuations of the two QDs follow a Lorentzian distribution in frequency with a full-width-at-half-maximum of $\Xi$. Assuming identical QDs, $\Gamma^*_{i,j} = \pi\Delta\nu_{H}$, $\Xi = 2\Delta\nu_{S}$, with $\Delta\nu_{H}$ and $\Delta\nu_{S}$ representing the homogeneous and inhomogeneous linewidth broadening of each QD, respectively. In the case of $\delta t = 0$ and $\Delta = 0$, Eq.\ \ref{eq:m3} simplifies to Eq.\ \ref{Vtheory}.  

Utilising Eq.\ \ref{eq:m3}, we estimate the effects of both pure dephasing and spectral fluctuations on the two-QD HOM experiments -- we determine $\Gamma^*_{i,j}=34\pm25\ \mathrm{MHz}$ and $\Xi = 2\times (34\pm 15)\ \mathrm{MHz}$ (Supplementary Note 2.B). In the time domain, the exciton dephasing rate corresponds to a $T_{\mathrm{2}}$-like time of $29 \pm 21\ \mathrm{ns}$; the spectral fluctuations correspond to a $T_{\mathrm{2}}^*$-like time of $9.1 \pm 3.9\ \mathrm{ns}$. 

\subsection{Effects of temporal post-selection and spectral filtering} \label{time-binning}

In the data analysis process, the HOM visibilities are calculated in a rigorous way: for every HOM peak, the coincidence events in a whole repetition pulse-period are considered, i.e.\ $T_{\mathrm{bin}} = T_{\mathrm{period}}$. Narrowing the evaluation time-window $T_{\mathrm{bin}}$ leads to temporal post-selections [Extended Data Fig.\ \ref{fig:timebin}(a)]. The time dependence of the intensity correlation function in the HOM experiment can be calculated using our theoretical model. This allows the dependence of the two-QD HOM visibility on the width of $T_{\mathrm{bin}}$ to be calculated [Extended Data Fig.\ \ref{fig:timebin}(b)]. 
As $T_{\mathrm{bin}}$ approaches $T_{\mathrm{period}}$ we find $\mathcal{V}_\mathrm{calc} = 93\%$, equivalent to the results from Eq.\ \ref{Vtheory} (a consequence of $T_{\mathrm{bin}}\gg\tau_{1,2}$). As $T_{\mathrm{bin}}$ decreases, we observe first almost no change in $\mathcal{V}_\mathrm{calc} $ until $T_{\mathrm{bin}}\sim20\ \tau_{r}$ ($\simeq 5.2$ ns). Reducing $T_{\mathrm{bin}}$ further results in a sharp increase in the calculated visibility. For example, at $T_{\mathrm{bin}} = 2\ \tau_{r}$ we find $\mathcal{V}_\mathrm{calc} = 98\%$. In this case, more than 50$\%$ of the coincidence counts are rejected by the post-selection. 

We also perform the analysis on the experimental results of the two-QD HOM measurement as a function of $T_{\mathrm{bin}}$. An example is shown in Extended Data Fig.\ \ref{fig:timebin}(c) for the remote HOM experiment using QD1 and QD2. Similar to the theoretical prediction, we see a decrease in $\mathcal{V}_{\mathrm{exp}}$ as $T_{\mathrm{bin}}$ increases. The dependence of the two-QD HOM visibility on the width of time-window $T_{\mathrm{bin}}$ matches the theoretical result rather well.

Sending QD photons through a narrow-bandwidth spectral filter has a similar effect compared to temporal post-selections. To simulate the spectral filtering effect, we consider again two identical QDs with radiative rates of $\Gamma$. Besides, we assume three different settings for the noise level in the QDs, accounting for 20\% (blue), 40\% (orange) and 60\% (yellow) of the total radiative rate, respectively. 

On the one hand, a narrow spectral-filter reduces the relative ratio of dephasing $\Gamma_{\mathrm{sum}}^*$ and spectral fluctuations $\pi\Xi$ with respect to the radiative rate $\Gamma$. Here, $\Gamma_{\mathrm{sum}}^*= 2\pi\Delta\nu_{H}$ is the QDs' total dephasing rate. We calculate how the ratio between the noise-related broadening and the intrinsic QD spectrum (defined as $\mathcal{A}_{\mathrm{noise}}$) varies after passing through a spectral filter (bandwidth $\Delta v_{\mathrm{fil}}$). Based on the effective noise level $\mathcal{A}_{\mathrm{noise}}$, we estimated the two-QD HOM visibility $\mathcal{V}_{\mathrm{est}}$ after spectral filtering, see Extended Data Fig.\ \ref{fig:filter}(a). From the simulation, only when $\Delta v_{\mathrm{fil}}$ becomes small with respect to the QD intrinsic linewidth, the filtering effect becomes prominent. $\mathcal{V}_{\mathrm{est}}$ rises steeply when $\Delta v_{\mathrm{fil}}$ is reduced to about five times of the QD intrinsic linewidth. 

On the other hand, the narrow filter can reject a significant part of the QD photons. In Extended Data Fig.\ \ref{fig:filter}(b), we calculate the probability $\eta_{\mathrm{fil}}$ with which the QD photons are not removed by the spectral filter (that is, transmit through the filter). $\eta_{\mathrm{fil}}$ decreases sharply, corresponding to strongly reduced count-rates, when the filter has a comparable bandwidth with respect to the QDs linewidth, e.g.\ $\Delta v_{\mathrm{fil}}(2\pi\tau_{\mathrm{r}})< 5$. Taking $\eta_{\mathrm{fil}} = 0.5$ as a limit, we plot the corresponding filtering effect on the two-QD HOM visibility in black circles in (a). Sacrificing half of the photons, we estimate that the two-QD HOM visibility can be boosted from 63\% to 77\% (yellow), from 71.4\% to 83\% (orange), and from 83\% to 91\% (blue) for the three noise levels. Sacrificing more than 50\% efficiency makes little sense in practical experiments, in particular for scaling to large photon numbers -- in this limit, a single QD-source creates more identical photons than multiple QD-sources.

Using the same noise-level settings, we investigate the effect of temporal post-selections. In Extended Data Fig.\ \ref{fig:filter}(c), we plot the post-selected two-QD HOM visibility $\mathcal{V}_{\mathrm{post}}$ as a function of $T_{\mathrm{bin}}$. The calculation is performed in a similar way compared to Extended Data Fig.\ \ref{fig:timebin}(b) -- the shape of delay dependence of the two-QD HOM interference is calculated, then a evaluation time-window is applied to the calculated shape yielding $\mathcal{V}_{\mathrm{post}}$. In Extended Data Fig.\ \ref{fig:filter}(d), we calculated the efficiency $\eta_{\mathrm{post}}$ with which the QD photons survive the temporal post-selection. At small $T_{\mathrm{bin}}$, $\mathcal{V}_{\mathrm{post}}$ increases and $\eta_{\mathrm{post}}$ decreases sharply. Taking $\eta_{\mathrm{post}} = 0.5$ as a limit (indicated as the black circles in (c)), we determine how much $\mathcal{V}_{\mathrm{post}}$ benefits from the temporal post-selection. For 60$\%$, 40$\%$, and 20$\%$ noise-levels, temporal post-selection can increase $\mathcal{V}_{\mathrm{post}}$ by (relative percentages) 29$\%$, 20$\%$ and 10$\%$, respectively, when sacrificing half of the coincidence events. The effect of temporal post-selection weakens as the noise-level becomes low. This highlights the challenge of achieving near-unity two-QD HOM visibility. Even with extreme post-selection ($\eta_{\mathrm{post}} = 0.5$), an unfiltered visibility of at least $86\%$ is required for a post-selected visibility to equal $93\%$ as reported in this work.

\subsection{Photon-photon entanglement using separate quantum dot sources}\label{optical}

Employing coherent photons from separate QDs, the optical CNOT circuit can produce a Bell state, $\ket{\Psi^-}$. In this section, we outline the analysis of this output entangled state. More details of the CNOT circuit are described in Supplementary Note 3.

The output entanglement is analysed by quantum state tomography consisting of $9\times4$ coincidence measurements in a combination of $\ket{H},\ket{V},\ket{+},\ket{-},\ket{R},\ket{L}$, where $\ket{L}= (\ket{H}+i\ket{V})/\sqrt{2}$ and $\ket{R}=(\ket{H}-i\ket{V})/\sqrt{2}$. As before, the coincidence counts of each measurement are calculated as the area under the central peak using $T_{\mathrm{bin}} = 13\ \mathrm{ns}$. The 36 projection bases $\ket{\psi_\nu}$ and their corresponding probabilities $S_{\nu}$ are depicted in Extended Data Fig.\ \ref{fig:sv_cnot}. The density matrix $\hat{\rho}$ is reconstructed based on these projection measurements using maximum likelihood estimation as detailed in Ref.\ \onlinecite{James2001,Altepeter2005}. 

From the reconstructed $\hat{\rho}$ we calculate the entanglement fidelity, which measures the overlaps between the experimentally generated state and the ideal state (density matrix $\hat{\rho}_{\mathrm{ideal}}$),
\begin{equation*}
  \mathcal{F}(\hat{\rho}_{\mathrm{ideal}},\hat{\rho}) = \Big(\mathrm{Tr}\big[\sqrt{\sqrt{\hat{\rho}_{\mathrm{ideal}}}\hat{\rho}\sqrt{\hat{\rho}_{\mathrm{ideal}}}}\big]\Big)^2.
\end{equation*}
For $\ket{\Psi^-}$, it simplifies to: 
\begin{equation}\label{eq:fcnotcal}
    \mathcal{F}_{\mathrm{\Psi}^-} = \bra{\Psi^-}\hat{\rho}\ket{\Psi^-}.
\end{equation}
Furthermore, the concurrence and the linear entropy are determined. The concurrence characterises the coherence properties of a quantum state. It is defined as:
 \begin{equation}\label{eq:crho}
     C(\hat{\rho}) = \mathrm{max}(0,\lambda_1-\lambda_2-\lambda_3-\lambda_4),
 \end{equation}
where $\lambda_{1}...\lambda_4$ represent the eigenvalues of the
product of the state $\rho$ and its spin-flipped counterpart $\tilde{\rho}$. Among them, $\lambda_{1}$ is the maximal eigenvalue. The spin-flip operation on the state $\rho$ reads: $\tilde{\rho} = (\sigma_y\otimes\sigma_y)\rho^*(\sigma_y\otimes\sigma_y$). The linear entropy $S_{\mathrm{L}}$ quantifies the degree of mixture in quantum states. $S_{\mathrm{L}}$ ranges from zero for pure states to one for completely mixed states.
 \begin{equation}\label{eq:linen}
     S_{\mathrm{L}}(\hat{\rho}) = \frac{4}{3}[1-\mathrm{Tr}(\hat{\rho}^2)].
 \end{equation}
Substituting the reconstructed density matrix $\hat{\rho}$ into Eqs.\ (\ref{eq:fcnotcal}-\ref{eq:linen}), we obtain $\mathcal{F}_{\mathrm{\Psi}^-} = (85.02\pm0.97)\%$, $C = (74.67\pm 1.93)\%$ and  $S_{\mathrm{L}} = (34.04\pm 1.94)\%$. The error margins are deduced from Monte-Carlo simulations assuming errors in the coincidence counts stemming from Poissonian statistics\cite{Altepeter2005}. 
The entanglement fidelity can be also estimated using the following equation, which takes into account six coincidence probabilities\cite{White2007},
\begin{equation}\label{eq:fcnot}
    \mathcal{F}_{\mathrm{\Psi}^-} = \frac{2-(S_{HH}+S_{VV}+S_{--}+S_{++}+S_{RR}+S_{LL})}{2}
\end{equation}
We can perform a consistence check using Eq.\ \ref{eq:fcnot}. Inserting the values from Extended Data Fig.\ \ref{fig:sv_cnot}, we arrive at $\mathcal{F}_{\mathrm{\Psi}^-} = 85.58\%$. This value lies within the 1$\sigma$-error margin of the result estimated with density-matrix reconstruction. 

The deviation from the ideal $\ket{\Psi^-}$ state likely originates from both setup imperfections, for instance the inaccurate rotation of wave-plates, and the remaining imperfections in the two-QD photons due to the non-perfect interference visibility and small values of $g^{(2)}(0)$. Assuming $g^{(2)}(0) = 0$ for both QDs and a perfect CNOT setup, we estimate the fidelity of entanglement to be $90.2\%$ based on the measured two-QD HOM visibility ($\mathcal{V} = 93.0\%$) using Eq.\ \ref{eq:fcnot}. The CNOT process and the produced entanglement should benefit from a boost of the two-QD HOM visibility when Purcell enhancements are introduced to the separate GaAs QD systems. Employing integrated photonic circuits can further reduce the setup imperfections.

\section*{Data Availability}
The raw data that support the findings of this study are available at \url{https://doi.org/10.5281/zenodo.6371310} and from the corresponding author upon reasonable request.
\section*{Code Availability}
The codes that have been used for this study are available from the corresponding author upon reasonable request.

\onecolumngrid
\clearpage

\section*{Extended Data}
\renewcommand{\figurename}{Extended Data Fig.}
\setcounter{figure}{0}

\begin{figure*}[h]
\centering
\includegraphics[width= 1 \columnwidth]{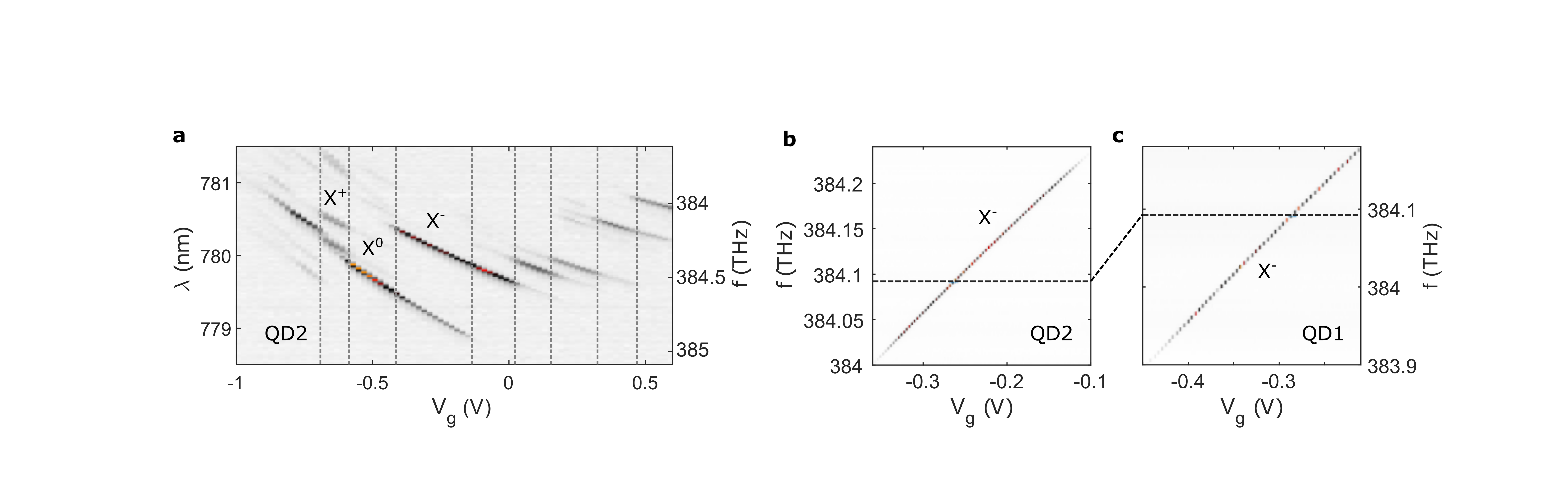}
\caption{\label{fig:plrf}{\bf Photoluminescence and resonance fluorescence charge plateaus. (a)} Photoluminescence from QD2 as a function of the externally applied gate voltage, $V_g$. The three excitons which can be resonantly excited are labelled: the positive trion $\mathrm{X}^{+}$, the neutral exciton $\mathrm{X}^{0}$ and the negative trion $\mathrm{X}^{-}$. \textbf{(b)} Resonance fluorescence on $\mathrm{X}^{-}$ from QD2. Resonance fluorescence is mapped out by scanning both the laser frequency and the gate voltage. The dashed line indicates the frequency at which all the experiments on QD2 are performed. \textbf{(c)} Resonance fluorescence on $\mathrm{X}^{-}$ from QD1. The dashed line represents the same frequency as in (b).}
\end{figure*}
\clearpage
\begin{figure*}[h]
\centering
\includegraphics[width= 1 \columnwidth]{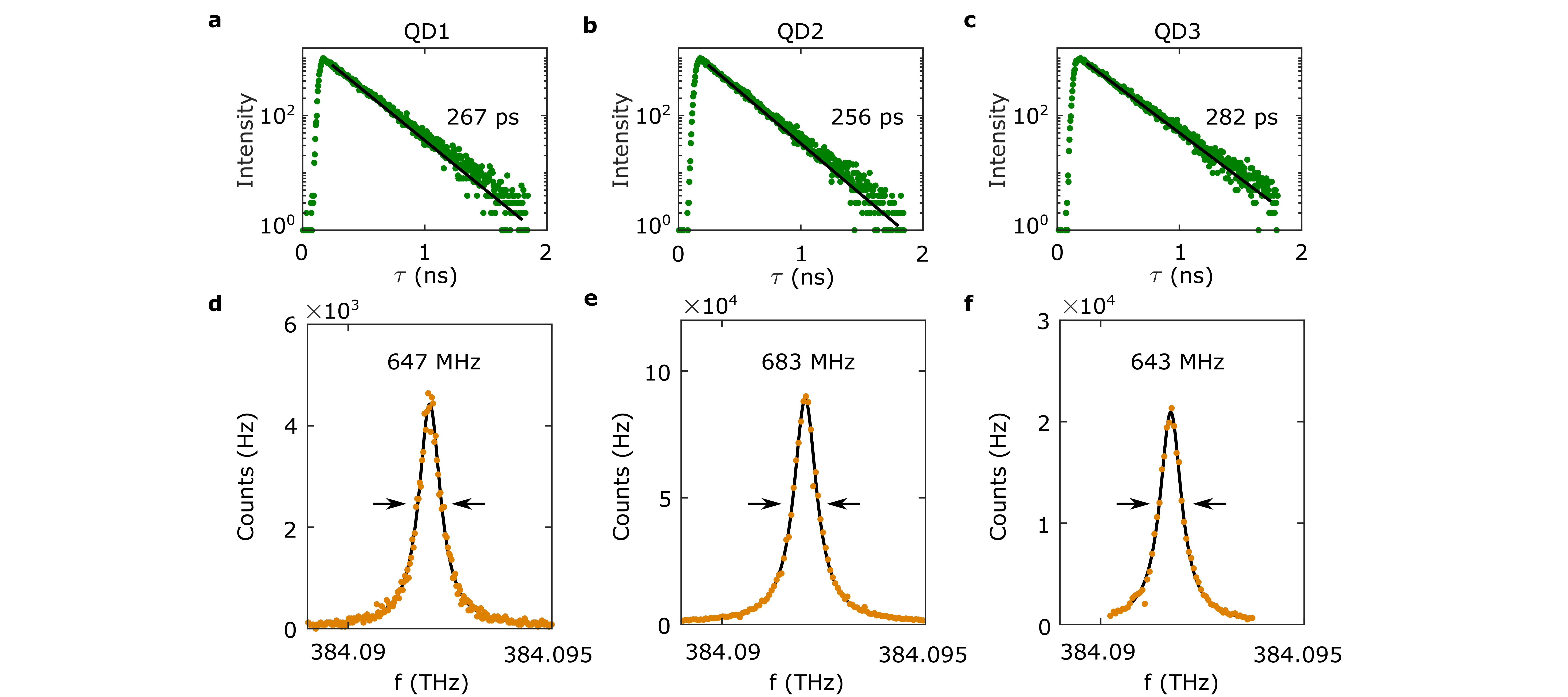}
\caption{\label{fig:l_l}{\bf Lifetimes and linewidths of the $\mathrm{X}^{-}$ in three quantum dots. (a-c)} Time-resolved resonance fluorescence from QD 1 - 3 under resonant pulsed excitation. The resonance fluorescence intensity of each QD follows an exponential decay. From the fits (black curves), the radiative decay rates are extracted as $\Gamma_1 = 3.75\ \mathrm{GHz}$, $\Gamma_2 = 3.91\ \mathrm{GHz}$ and $\Gamma_3 = 3.54\ \mathrm{GHz}$. The corresponding lifetimes are displayed next to the exponential fits. The radiative lifetime of GaAs QDs is typically\cite{Gurioli2019} around $250\ \mathrm{ps}$. In our sample, the average lifetime is around $300\ \mathrm{ps}$ -- there is no Purcell enhancement. \textbf{(d-f)} Spectrum of the resonance fluorescence obtained by slowly scanning a narrow-bandwidth continuous-wave (CW) laser across the $\mathrm{X}^{-}$. The typical measurement time is 5 - 10 minutes -- the linewidth probes the noise over a huge frequency bandwidth. The measured linewidths (values are displayed next to the fits) are very close to the lifetime limits.}
\end{figure*}
\clearpage
\begin{figure*}[h]
\includegraphics[width=1\columnwidth]{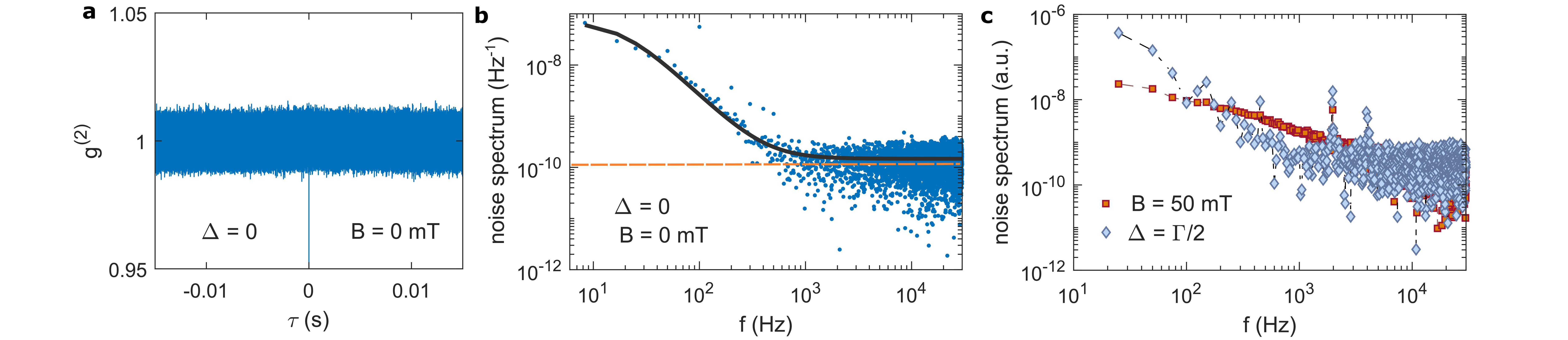}
\caption{\label{fig:nspectrum}\textbf{The noise in GaAs quantum dots. (a) }Auto-correlation measurements of X$^{-}$ in a GaAs QD in the diode heterostructure under continuous-wave (CW) excitation. The CW laser excites the QD resonantly. The auto-correlation $g^{(2)}$ is normalised by the mathematical expectation\cite{Zhai2020} based on the photon count-rates and the integration time. The $g^{(2)}$ is flat, a feature showing the absence of the blinking. \textbf{(b)} Noise spectrum of a GaAs QD under resonant excitation. Like (a), a narrow-bandwidth laser is placed at the exact resonance of X$^{-}$. The noise is resolved as a function of frequency f. The black curve represents a Lorentzian fit to the noise profile. The orange dashed line represents the shot-noise level. The bandwidth of the Lorentzian is extracted to be 19 Hz (half-width-at-half-maximum), showing that the environmental noise is concentrated at low frequencies. The noise spectrum at higher frequencies, e.g.\ 10$^4$-10$^6$ Hz, remains small and mostly flat. \textbf{(c)} Noise spectrum of a GaAs QD when the sensitivity of either charge noise or spin noise is increased. The light-blue curve represents the condition when the laser is detuned by half of the QD linewidth, $\Delta = \Gamma/2$. In this case, low-frequency ($<$10$^2$ Hz) charge noise is enhanced compared to (b). The orange curve represents the condition where the QD is placed in a small magnetic field $B = 50$ mT, and the laser frequency is placed in the centre of two Zeeman-split peaks. The low-frequency noise becomes less, but the noise within 10$^2$--10$^4$ Hz range is enhanced, which is likely the spin noise.}
\end{figure*}
\clearpage

\begin{figure}
\centering
\includegraphics[width= 1 \columnwidth]{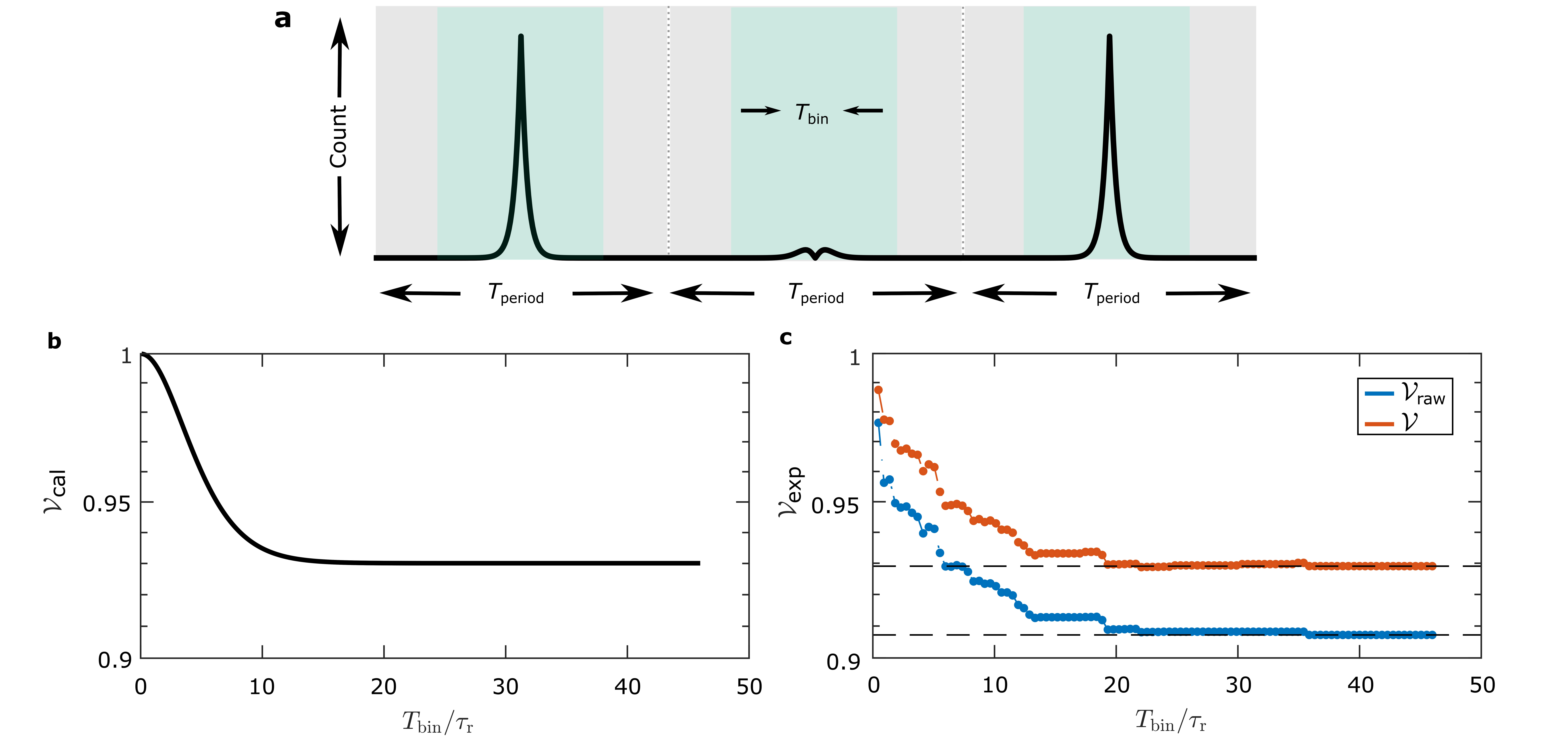}
\caption{\label{fig:timebin}{\bf Remote interference visibility as a function of evaluation time-window.} \textbf{(a)} A sketch showing the evaluation time-binning window in the HOM data analysis. The pulsed excitation laser has a repetition period of $T_{\mathrm{period}} = 13\ \mathrm{ns}$, which corresponds to $\sim 50\ \tau_{r}$. Every $T_{\mathrm{period}}$ contains one HOM peak, and the time-binning window (with a width of $T_{\mathrm{bin}}$) is centred around each HOM peak. In \textbf{(b,c)} we reduce the width of the time-window $T_{\mathrm{bin}}$ from $T_{\mathrm{period}}$ to close to zero, and calculate/extract the theoretical/measured value of the two-QD HOM visibility. In (b), we apply the time-binning window to a theoretical delay dependence two-photon interference $\mathcal{G}^{(2)}(\tau)$. This $\mathcal{G}^{(2)}(\tau)$ is calculated using Eq.\ 19 in Supplementary Information, with parameters $\Gamma_1 = 3.75\ \mathrm{GHz}$, $\Gamma_2 = 3.91\ \mathrm{GHz}$, $\Gamma^* = 34\ \mathrm{MHz}$, $\Xi =2\times 34\ \mathrm{MHz}$, $\delta t = 0$ and $\Delta = 0$. The calculated two-QD HOM visibility $\mathcal{V}_{\mathrm{cal}}$ drops to $93\%$ at $T_{\mathrm{bin}} = 20\ \tau_{r}$ and levels off. In (c), the measured two-QD HOM visibility (QD1 and QD2, $\delta t = 0$, $\Delta = 0$) is shown as a function of the normalised width of the time-window. The visibility can be effectively increased to $\mathcal{V}_{\mathrm{exp}}\sim 98\%$ when $T_{\mathrm{bin}}$ is comparable to the QD lifetime. At large time-windows ($T_{\mathrm{bin}} \rightarrow T_{\mathrm{period}}$), we determine the real two-QD HOM visibility, $\mathcal{V}= 93\%$: in this limit, no temporal post-selection is included.}
\end{figure}
\clearpage

\begin{figure*}[p]
\includegraphics[width=1\columnwidth]{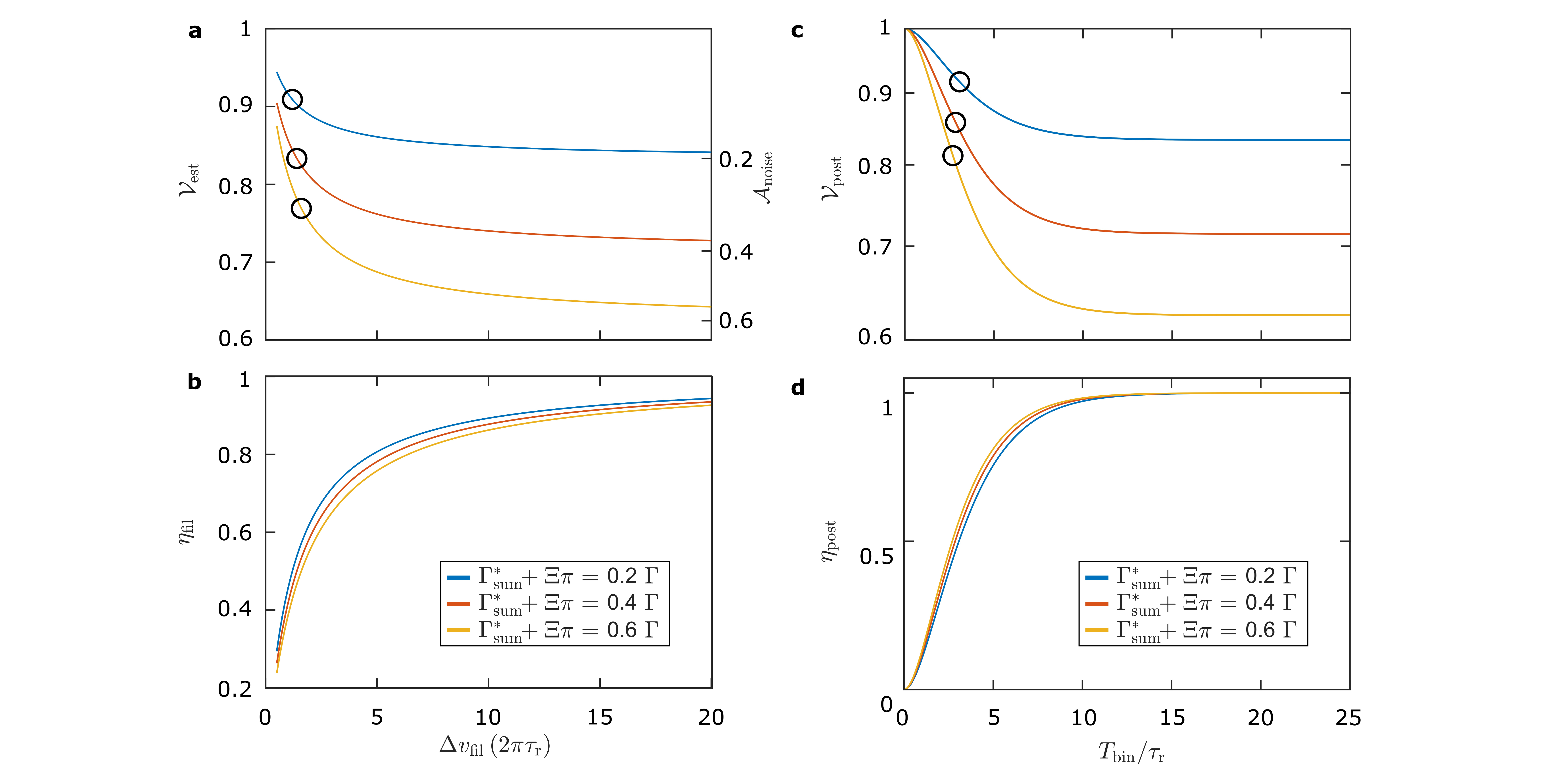}
\caption{\label{fig:filter}\textbf{Effects of spectral filtering and temporal post-selection on two-photon interference from remote quantum dots.} \textbf{(a)} The effect of spectral filtering as a function of the filter bandwidth $\Delta v_{\mathrm{fil}}$. We study here a filter with a narrow bandwidth and assume a Lorentzian transmission function (e.g.\ for an etalon). $\Delta v_{\mathrm{fil}}$ is normalised by the QD's radiative rate $1/(2\pi\tau_{\mathrm{r}})$ (assuming identical QDs). $\mathcal{A}_{\mathrm{noise}}$ is an indicator of the spectral filtering effect. It is defined as the ratio between the noise-related linewidth broadening and the QD's intrinsic linewidth. Different coloured lines represent different levels of noise, characterised by ($\Gamma_{\mathrm{sum}}^*+\Xi\pi$)$/\Gamma$. Both the intrinsic and the noise part of the QD spectrum experience spectral filtering effects, but their ratio $\mathcal{A}_{\mathrm{noise}}$ decreases as the filter narrows. However, this reduction in $\mathcal{A}_{\mathrm{noise}}$ only becomes apparent when the filter is narrow, e.g.\ $\Delta v_{\mathrm{fil}}(2\pi\tau_{\mathrm{r}})< 5$. \textbf{(b)} The effect of spectral filtering on the photon counts. $\eta_{\mathrm{fil}}$ represents the percentage of photons exiting a spectral filter compared to the photons before filtering. Here, the peak transmission of the filter is set to unity and the filter is exactly centred at the QD resonance, an idealised situation. \textbf{(c)} The effect of temporal post-selection on two-QD HOM interference as a function of the width of evaluation time-window $T_{\mathrm{bin}}$. Performing temporal post-selection with a narrow $T_{\mathrm{bin}}$ leads to an increase in two-QD HOM visibility $\mathcal{V}_{\mathrm{post}}$ at the expense of coincidence count-rates. Here, the line colours again represent the different noise conditions. Similar to spectral filtering, the effect of post-selection only becomes prominent at small $T_{\mathrm{bin}}$. \textbf{(d)} The effect of temporal post-selection on coincidence counts. $\eta_{\mathrm{post}}$ is defined as the ratio between the total coincidence events after temporal post-selection compared to the no post-selection case.} 
\end{figure*}
\clearpage
\begin{figure}[p]
\centering
\includegraphics[width= 1 \columnwidth]{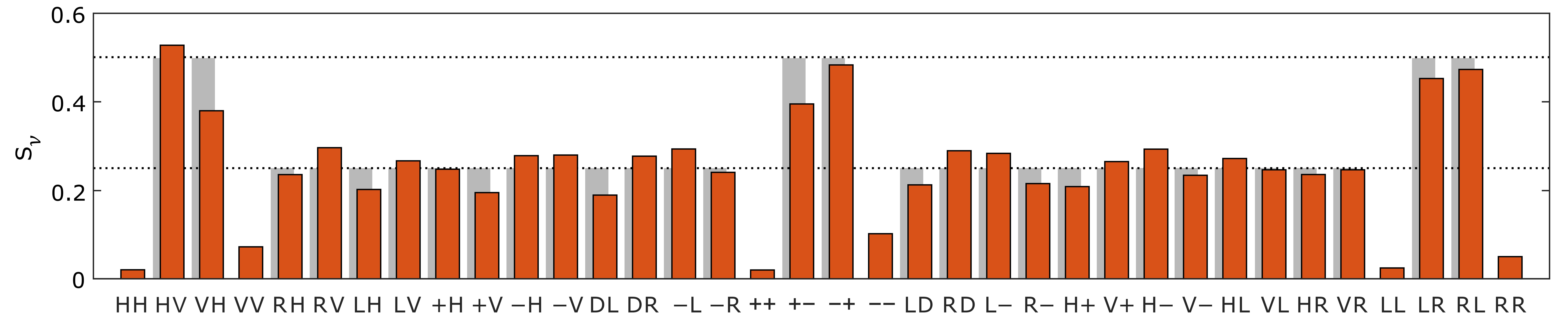}
\caption{\label{fig:sv_cnot}{\bf Projection outcomes of $\ket{\Psi^-}$ in quantum state tomography.} The probabilities $S_{\nu}$ are calculated by adding up the coincidence counts in the central peak and normalising the sum to the overall counts in each set of four coincidence measurements. Here, 36 possible $\nu$-states are listed in the x-axis. The two dashed lines indicate the $S_{\nu} = 0.25$ and $S_{\nu}=0.5$ levels. The light grey background represents the projection probabilities for the ideal $\ket{\Psi^-}$ state.}
\end{figure}


\clearpage

\begin{thebibliography}{10}
\expandafter\ifx\csname url\endcsname\relax
  \def\url#1{\texttt{#1}}\fi
\expandafter\ifx\csname urlprefix\endcsname\relax\def\urlprefix{URL }\fi
\providecommand{\bibinfo}[2]{#2}
\providecommand{\eprint}[2][]{\url{#2}}

\bibitem{Sangouard2011}
\bibinfo{author}{Sangouard, N.}, \bibinfo{author}{Simon, C.},
  \bibinfo{author}{de~Riedmatten, H.} \& \bibinfo{author}{Gisin, N.}
\newblock \bibinfo{title}{Quantum repeaters based on atomic ensembles and
  linear optics}.
\newblock \emph{\bibinfo{journal}{Rev. Mod. Phys.}}
  \textbf{\bibinfo{volume}{83}}, \bibinfo{pages}{33--80}
  (\bibinfo{year}{2011}).

\bibitem{Yin2016}
\bibinfo{author}{Yin, H.-L.} \emph{et~al.}
\newblock \bibinfo{title}{Measurement-device-independent quantum key
  distribution over a 404 km optical fiber}.
\newblock \emph{\bibinfo{journal}{Phys. Rev. Lett.}}
  \textbf{\bibinfo{volume}{117}}, \bibinfo{pages}{190501}
  (\bibinfo{year}{2016}).

\bibitem{Wang2019a}
\bibinfo{author}{Wang, H.} \emph{et~al.}
\newblock \bibinfo{title}{Boson sampling with 20 input photons and a 60-mode
  interferometer in a $1{0}^{14}$-dimensional {H}ilbert space}.
\newblock \emph{\bibinfo{journal}{Phys. Rev. Lett.}}
  \textbf{\bibinfo{volume}{123}}, \bibinfo{pages}{250503}
  (\bibinfo{year}{2019}).

\bibitem{Qiang2018}
\bibinfo{author}{Qiang, X.} \emph{et~al.}
\newblock \bibinfo{title}{Large-scale silicon quantum photonics implementing
  arbitrary two-qubit processing}.
\newblock \emph{\bibinfo{journal}{Nat. Photonics}}
  \textbf{\bibinfo{volume}{12}}, \bibinfo{pages}{534--539}
  (\bibinfo{year}{2018}).

\bibitem{Strauf2007}
\bibinfo{author}{Strauf, S.} \emph{et~al.}
\newblock \bibinfo{title}{High-frequency single-photon source with polarization
  control}.
\newblock \emph{\bibinfo{journal}{Nat. Photonics}}
  \textbf{\bibinfo{volume}{1}}, \bibinfo{pages}{704--708}
  (\bibinfo{year}{2007}).

\bibitem{Senellart2017}
\bibinfo{author}{Senellart, P.}, \bibinfo{author}{Solomon, G.} \&
  \bibinfo{author}{White, A.}
\newblock \bibinfo{title}{High-performance semiconductor quantum-dot
  single-photon sources}.
\newblock \emph{\bibinfo{journal}{Nat. Nanotechnol.}}
  \textbf{\bibinfo{volume}{12}}, \bibinfo{pages}{1026--1039}
  (\bibinfo{year}{2017}).

\bibitem{Liu2018}
\bibinfo{author}{Liu, F.} \emph{et~al.}
\newblock \bibinfo{title}{High {P}urcell factor generation of indistinguishable
  on-chip single photons}.
\newblock \emph{\bibinfo{journal}{Nat. Nanotechnol.}}
  \textbf{\bibinfo{volume}{13}}, \bibinfo{pages}{835--840}
  (\bibinfo{year}{2018}).

\bibitem{Uppu2021}
\bibinfo{author}{Uppu, R.}, \bibinfo{author}{Midolo, L.},
  \bibinfo{author}{Zhou, X.}, \bibinfo{author}{Carolan, J.} \&
  \bibinfo{author}{Lodahl, P.}
\newblock \bibinfo{title}{Quantum-dot-based deterministic photon–emitter
  interfaces for scalable photonic quantum technology}.
\newblock \emph{\bibinfo{journal}{Nat. Nanotechnol.}}  (\bibinfo{year}{2021}).

\bibitem{Tomm2021}
\bibinfo{author}{Tomm, N.} \emph{et~al.}
\newblock \bibinfo{title}{{A bright and fast source of coherent single
  photons}}.
\newblock \emph{\bibinfo{journal}{Nat. Nanotechnol.}}
  \textbf{\bibinfo{volume}{16}}, \bibinfo{pages}{399--403}
  (\bibinfo{year}{2021}).

\bibitem{Reindl2017}
\bibinfo{author}{Reindl, M.} \emph{et~al.}
\newblock \bibinfo{title}{Phonon-assisted two-photon interference from remote
  quantum emitters}.
\newblock \emph{\bibinfo{journal}{Nano Lett.}} \textbf{\bibinfo{volume}{17}},
  \bibinfo{pages}{4090--4095} (\bibinfo{year}{2017}).

\bibitem{Weber2018}
\bibinfo{author}{Weber, J.~H.} \emph{et~al.}
\newblock \bibinfo{title}{Two-photon interference in the telecom {C}-band after
  frequency conversion of photons from remote quantum emitters}.
\newblock \emph{\bibinfo{journal}{Nat. Nanotechnol.}}
  \textbf{\bibinfo{volume}{14}}, \bibinfo{pages}{23--26}
  (\bibinfo{year}{2018}).

\bibitem{Llewellyn2020}
\bibinfo{author}{Llewellyn, D.} \emph{et~al.}
\newblock \bibinfo{title}{Chip-to-chip quantum teleportation and multi-photon
  entanglement in silicon}.
\newblock \emph{\bibinfo{journal}{Nat. Phys.}} \textbf{\bibinfo{volume}{16}},
  \bibinfo{pages}{148--153} (\bibinfo{year}{2020}).

\bibitem{He2013_2}
\bibinfo{author}{He, Y.-M.} \emph{et~al.}
\newblock \bibinfo{title}{On-demand semiconductor single-photon source with
  near-unity indistinguishability}.
\newblock \emph{\bibinfo{journal}{Nat. Nanotechnol.}}
  \textbf{\bibinfo{volume}{8}}, \bibinfo{pages}{213--217}
  (\bibinfo{year}{2013}).

\bibitem{Basset2021}
\bibinfo{author}{Basset, F.~B.} \emph{et~al.}
\newblock \bibinfo{title}{Quantum key distribution with entangled photons
  generated on demand by a quantum dot}.
\newblock \emph{\bibinfo{journal}{Sci. Adv.}} \textbf{\bibinfo{volume}{7}},
  \bibinfo{pages}{eabe6379} (\bibinfo{year}{2021}).

\bibitem{Grim2019}
\bibinfo{author}{Grim, J.~Q.} \emph{et~al.}
\newblock \bibinfo{title}{Scalable in operando strain tuning in nanophotonic
  waveguides enabling three-quantum-dot superradiance}.
\newblock \emph{\bibinfo{journal}{Nat. Mater.}} \textbf{\bibinfo{volume}{18}},
  \bibinfo{pages}{963--969} (\bibinfo{year}{2019}).

\bibitem{Kolodynski2020}
\bibinfo{author}{Kołodyński, J.} \emph{et~al.}
\newblock \bibinfo{title}{Device-independent quantum key distribution with
  single-photon sources}.
\newblock \emph{\bibinfo{journal}{Quantum}} \textbf{\bibinfo{volume}{4}},
  \bibinfo{pages}{260} (\bibinfo{year}{2020}).

\bibitem{Patel2010}
\bibinfo{author}{Patel, R.~B.} \emph{et~al.}
\newblock \bibinfo{title}{Two-photon interference of the emission from
  electrically tunable remote quantum dots}.
\newblock \emph{\bibinfo{journal}{Nat. Photonics}}
  \textbf{\bibinfo{volume}{4}}, \bibinfo{pages}{632--635}
  (\bibinfo{year}{2010}).

\bibitem{He2013}
\bibinfo{author}{He, Y.} \emph{et~al.}
\newblock \bibinfo{title}{Indistinguishable tunable single photons emitted by
  spin-flip raman transitions in {I}n{G}a{A}s quantum dots}.
\newblock \emph{\bibinfo{journal}{Phys. Rev. Lett.}}
  \textbf{\bibinfo{volume}{111}}, \bibinfo{pages}{237403}
  (\bibinfo{year}{2013}).

\bibitem{Giesz2015}
\bibinfo{author}{Giesz, V.} \emph{et~al.}
\newblock \bibinfo{title}{Cavity-enhanced two-photon interference using remote
  quantum dot sources}.
\newblock \emph{\bibinfo{journal}{Phys. Rev. B}} \textbf{\bibinfo{volume}{92}},
  \bibinfo{pages}{161302} (\bibinfo{year}{2015}).

\bibitem{Zopf2018}
\bibinfo{author}{Zopf, M.} \emph{et~al.}
\newblock \bibinfo{title}{Frequency feedback for two-photon interference from
  separate quantum dots}.
\newblock \emph{\bibinfo{journal}{Phys. Rev. B}} \textbf{\bibinfo{volume}{98}},
  \bibinfo{pages}{161302} (\bibinfo{year}{2018}).

\bibitem{You2021}
\bibinfo{author}{You, X.} \emph{et~al.}
\newblock \bibinfo{title}{Quantum interference between independent solid-state
  single-photon sources separated by 300 km fiber}.
\newblock arXiv:\eprint{2106.15545 (2021)}.

\bibitem{Zhai2020}
\bibinfo{author}{Zhai, L.} \emph{et~al.}
\newblock \bibinfo{title}{Low-noise {GaAs} quantum dots for quantum photonics}.
\newblock \emph{\bibinfo{journal}{Nat. Commun.}} \textbf{\bibinfo{volume}{11}},
  \bibinfo{pages}{4745} (\bibinfo{year}{2020}).

\bibitem{Santori2002}
\bibinfo{author}{Santori, C.}, \bibinfo{author}{Fattal, D.},
  \bibinfo{author}{Vu{\v{c}}kovi{\'c}, J.}, \bibinfo{author}{Solomon, G.~S.} \&
  \bibinfo{author}{Yamamoto, Y.}
\newblock \bibinfo{title}{Indistinguishable photons from a single-photon
  device}.
\newblock \emph{\bibinfo{journal}{Nature}} \textbf{\bibinfo{volume}{419}},
  \bibinfo{pages}{594--597} (\bibinfo{year}{2002}).

\bibitem{Wang2016}
\bibinfo{author}{Wang, H.} \emph{et~al.}
\newblock \bibinfo{title}{Near-{T}ransform-limited single photons from an
  efficient solid-state quantum emitter}.
\newblock \emph{\bibinfo{journal}{Phys. Rev. Lett.}}
  \textbf{\bibinfo{volume}{116}}, \bibinfo{pages}{213601}
  (\bibinfo{year}{2016}).

\bibitem{Thoma2016}
\bibinfo{author}{Thoma, A.} \emph{et~al.}
\newblock \bibinfo{title}{Exploring dephasing of a solid-state quantum emitter
  via time- and temperature-dependent {Hong-Ou-Mandel} experiments}.
\newblock \emph{\bibinfo{journal}{Phys. Rev. Lett.}}
  \textbf{\bibinfo{volume}{116}}, \bibinfo{pages}{033601}
  (\bibinfo{year}{2016}).

\bibitem{Kuhlmann2013}
\bibinfo{author}{Kuhlmann, A.~V.} \emph{et~al.}
\newblock \bibinfo{title}{{Charge noise and spin noise in a semiconductor
  quantum device}}.
\newblock \emph{\bibinfo{journal}{Nat. Phys.}} \textbf{\bibinfo{volume}{9}},
  \bibinfo{pages}{570--575} (\bibinfo{year}{2013}).

\bibitem{Scholl2019}
\bibinfo{author}{Sch\"oll, E.} \emph{et~al.}
\newblock \bibinfo{title}{Resonance fluorescence of {G}a{A}s quantum dots with
  near-unity photon indistinguishability}.
\newblock \emph{\bibinfo{journal}{Nano Lett.}} \textbf{\bibinfo{volume}{19}},
  \bibinfo{pages}{2404--2410} (\bibinfo{year}{2019}).

\bibitem{Maunz2007}
\bibinfo{author}{Maunz, P.} \emph{et~al.}
\newblock \bibinfo{title}{Quantum interference of photon pairs from two remote
  trapped atomic ions}.
\newblock \emph{\bibinfo{journal}{Nat. Phys.}} \textbf{\bibinfo{volume}{3}},
  \bibinfo{pages}{538--541} (\bibinfo{year}{2007}).

\bibitem{Stephenson2020}
\bibinfo{author}{Stephenson, L.~J.} \emph{et~al.}
\newblock \bibinfo{title}{High-rate, high-fidelity entanglement of qubits
  across an elementary quantum network}.
\newblock \emph{\bibinfo{journal}{Phys. Rev. Lett.}}
  \textbf{\bibinfo{volume}{124}}, \bibinfo{pages}{110501}
  (\bibinfo{year}{2020}).

\bibitem{Beugnon2006}
\bibinfo{author}{Beugnon, J.} \emph{et~al.}
\newblock \bibinfo{title}{Quantum interference between two single photons
  emitted by independently trapped atoms}.
\newblock \emph{\bibinfo{journal}{Nature}} \textbf{\bibinfo{volume}{440}},
  \bibinfo{pages}{779–782} (\bibinfo{year}{2006}).

\bibitem{Stockill2017}
\bibinfo{author}{Stockill, R.} \emph{et~al.}
\newblock \bibinfo{title}{Phase-tuned entangled state generation between
  distant spin qubits}.
\newblock \emph{\bibinfo{journal}{Phys. Rev. Lett.}}
  \textbf{\bibinfo{volume}{119}}, \bibinfo{pages}{010503}
  (\bibinfo{year}{2017}).

\bibitem{Bernien2013}
\bibinfo{author}{Bernien, H.} \emph{et~al.}
\newblock \bibinfo{title}{Heralded entanglement between solid-state qubits
  separated by three metres}.
\newblock \emph{\bibinfo{journal}{Nature}} \textbf{\bibinfo{volume}{497}},
  \bibinfo{pages}{86–90} (\bibinfo{year}{2013}).

\bibitem{Humphreys2018}
\bibinfo{author}{Humphreys, P.~C.} \emph{et~al.}
\newblock \bibinfo{title}{Deterministic delivery of remote entanglement on a
  quantum network}.
\newblock \emph{\bibinfo{journal}{Nature}} \textbf{\bibinfo{volume}{558}},
  \bibinfo{pages}{268–273} (\bibinfo{year}{2018}).

\bibitem{Kambs2018}
\bibinfo{author}{Kambs, B.} \& \bibinfo{author}{Becher, C.}
\newblock \bibinfo{title}{{Limitations on the indistinguishability of photons
  from remote solid state sources}}.
\newblock \emph{\bibinfo{journal}{New J. Phys.}} \textbf{\bibinfo{volume}{20}},
  \bibinfo{pages}{115003} (\bibinfo{year}{2018}).

\bibitem{Kiesel2005}
\bibinfo{author}{Kiesel, N.}, \bibinfo{author}{Schmid, C.},
  \bibinfo{author}{Weber, U.}, \bibinfo{author}{Ursin, R.} \&
  \bibinfo{author}{Weinfurter, H.}
\newblock \bibinfo{title}{Linear optics controlled-phase gate made simple}.
\newblock \emph{\bibinfo{journal}{Phys. Rev. Lett.}}
  \textbf{\bibinfo{volume}{95}}, \bibinfo{pages}{210505}
  (\bibinfo{year}{2005}).

\bibitem{James2001}
\bibinfo{author}{James, D. F.~V.}, \bibinfo{author}{Kwiat, P.~G.},
  \bibinfo{author}{Munro, W.~J.} \& \bibinfo{author}{White, A.~G.}
\newblock \bibinfo{title}{Measurement of qubits}.
\newblock \emph{\bibinfo{journal}{Phys. Rev. A}} \textbf{\bibinfo{volume}{64}},
  \bibinfo{pages}{052312} (\bibinfo{year}{2001}).

\bibitem{Istrati2020}
\bibinfo{author}{Istrati, D.} \emph{et~al.}
\newblock \bibinfo{title}{Sequential generation of linear cluster states from a
  single photon emitter}.
\newblock \emph{\bibinfo{journal}{Nat. Commun.}} \textbf{\bibinfo{volume}{11}}
  (\bibinfo{year}{2020}).

\bibitem{Cogan2021}
\bibinfo{author}{Cogan, D.}, \bibinfo{author}{Su, Z.-E.},
  \bibinfo{author}{Kenneth, O.} \& \bibinfo{author}{Gershoni, D.}
\newblock \bibinfo{title}{A deterministic source of indistinguishable photons
  in a cluster state}.
\newblock arXiv:\eprint{2110.05908 (2021)}.

\bibitem{Wolters2017}
\bibinfo{author}{Wolters, J.} \emph{et~al.}
\newblock \bibinfo{title}{Simple atomic quantum memory suitable for
  semiconductor quantum dot single photons}.
\newblock \emph{\bibinfo{journal}{Phys. Rev. Lett.}}
  \textbf{\bibinfo{volume}{119}}, \bibinfo{pages}{060502}
  (\bibinfo{year}{2017}).

\bibitem{Nguyen2020}
\bibinfo{author}{Nguyen, G.} \emph{et~al.}
\newblock \bibinfo{title}{Influence of molecular beam effusion cell quality on
  optical and electrical properties of quantum dots and quantum wells}.
\newblock \emph{\bibinfo{journal}{J. Cryst. Growth}}
  \textbf{\bibinfo{volume}{550}}, \bibinfo{pages}{125884}
  (\bibinfo{year}{2020}).

\bibitem{Gurioli2019}
\bibinfo{author}{Gurioli, M.}, \bibinfo{author}{Wang, Z.},
  \bibinfo{author}{Rastelli, A.}, \bibinfo{author}{Kuroda, T.} \&
  \bibinfo{author}{Sanguinetti, S.}
\newblock \bibinfo{title}{Droplet epitaxy of semiconductor nanostructures for
  quantum photonic devices}.
\newblock \emph{\bibinfo{journal}{Nat. Mater.}} \textbf{\bibinfo{volume}{18}},
  \bibinfo{pages}{799--810} (\bibinfo{year}{2019}).

\bibitem{Heyn2009}
\bibinfo{author}{Heyn, C.} \emph{et~al.}
\newblock \bibinfo{title}{Highly uniform and strain-free {G}a{A}s quantum dots
  fabricated by filling of self-assembled nanoholes}.
\newblock \emph{\bibinfo{journal}{Appl. Phys. Lett.}}
  \textbf{\bibinfo{volume}{94}}, \bibinfo{pages}{183113}
  (\bibinfo{year}{2009}).

\bibitem{Mooney1990}
\bibinfo{author}{Mooney, P.}
\newblock \bibinfo{title}{Deep donor levels ({D}{X} centers) in {I}{I}{I}-{V}
  semiconductors}.
\newblock \emph{\bibinfo{journal}{J. Appl. Phys.}}
  \textbf{\bibinfo{volume}{67}}, \bibinfo{pages}{R1--R26}
  (\bibinfo{year}{1990}).

\bibitem{Warburton2013}
\bibinfo{author}{Warburton, R.~J.}
\newblock \bibinfo{title}{Single spins in self-assembled quantum dots}.
\newblock \emph{\bibinfo{journal}{Nat. Mater.}} \textbf{\bibinfo{volume}{12}},
  \bibinfo{pages}{483} (\bibinfo{year}{2013}).

\bibitem{Kuhlmann2013a}
\bibinfo{author}{Kuhlmann, A.~V.} \emph{et~al.}
\newblock \bibinfo{title}{{A dark-field microscope for background-free
  detection of resonance fluorescence from single semiconductor quantum dots
  operating in a set-and-forget mode}}.
\newblock \emph{\bibinfo{journal}{Rev. Sci. Instrum.}}
  \textbf{\bibinfo{volume}{84}}, \bibinfo{pages}{073905}
  (\bibinfo{year}{2013}).

\bibitem{Lobl2019}
\bibinfo{author}{L\"obl, M.~C.} \emph{et~al.}
\newblock \bibinfo{title}{Correlations between optical properties and
  {V}oronoi-cell area of quantum dots}.
\newblock \emph{\bibinfo{journal}{Phys. Rev. B}}
  \textbf{\bibinfo{volume}{100}}, \bibinfo{pages}{155402}
  (\bibinfo{year}{2019}).

\bibitem{Keil2017}
\bibinfo{author}{Keil, R.} \emph{et~al.}
\newblock \bibinfo{title}{Solid-state ensemble of highly entangled photon
  sources at rubidium atomic transitions}.
\newblock \emph{\bibinfo{journal}{Nat. Commun.}} \textbf{\bibinfo{volume}{8}},
  \bibinfo{pages}{15501} (\bibinfo{year}{2017}).

\bibitem{Zhai2020_2}
\bibinfo{author}{Zhai, L.} \emph{et~al.}
\newblock \bibinfo{title}{{Large-range frequency tuning of a narrow-linewidth
  quantum emitter}}.
\newblock \emph{\bibinfo{journal}{Appl. Phys. Lett.}}
  \textbf{\bibinfo{volume}{117}}, \bibinfo{pages}{083106}
  (\bibinfo{year}{2020}).

\bibitem{Fischer2017}
\bibinfo{author}{Fischer, K.~A.} \emph{et~al.}
\newblock \bibinfo{title}{Signatures of two-photon pulses from a quantum
  two-level system}.
\newblock \emph{\bibinfo{journal}{Nat. Phys.}} \textbf{\bibinfo{volume}{13}},
  \bibinfo{pages}{649–654} (\bibinfo{year}{2017}).

\bibitem{Altepeter2005}
\bibinfo{author}{Altepeter, J.}, \bibinfo{author}{Jeffrey, E.} \&
  \bibinfo{author}{Kwiat, P.}
\newblock \bibinfo{title}{Photonic state tomography}.
\newblock Advances In Atomic, Molecular, and Optical Physics,
  \bibinfo{pages}{105--159} (\bibinfo{publisher}{Academic Press},
  \bibinfo{year}{2005}).

\bibitem{White2007}
\bibinfo{author}{White, A.~G.} \emph{et~al.}
\newblock \bibinfo{title}{Measuring two-qubit gates}.
\newblock \emph{\bibinfo{journal}{J. Opt. Soc. Am. B}}
  \textbf{\bibinfo{volume}{24}}, \bibinfo{pages}{172} (\bibinfo{year}{2007}).

\end{thebibliography}
\end{document}